%% file: mnras_template.tex
\documentclass[fleqn,usenatbib]{mnras}

\usepackage{newtxtext,newtxmath}
\usepackage{supertabular}

\usepackage[T1]{fontenc}

\DeclareRobustCommand{\VAN}[3]{#2}
\let\VANthebibliography\thebibliography
\def\thebibliography{\DeclareRobustCommand{\VAN}[3]{##3}\VANthebibliography}

\usepackage{subcaption}
\usepackage{longtable}
\usepackage{supertabular}
\usepackage{booktabs}
\usepackage{graphicx}	
\usepackage{amsmath}	
\usepackage[dvipsnames]{xcolor}

\newcommand{\msol}{$M_{\odot}$}

\newcommand{\resub}[1]{#1}

\title[The fate of rotating massive stars across cosmic times]{The fate of rotating massive stars across cosmic times}

\author[Hirschi et al.]{R. Hirschi$^{1,2}$\thanks{E-mail: r.hirschi@keele.ac.uk}, 
K. Goodman$^{1}$,
G. Meynet$^{3}$,
A. Maeder$^{3}$,
S. Ekstr\"om$^{3}$,
P. Eggenberger$^{3}$,
C. Georgy$^{3}$,
\newauthor 
Y. Sibony$^{3}$
N. Yusof$^{4}$,
S. Martinet$^{5}$,
Vishnu Varma$^{1}$
and 
K. Nomoto$^{2}$
\\
$^{1}$Astrophysics Research Centre, Lennard-Jones Laboratories, Keele University, Keele ST5 5BG, UK\\
$^{2}$Kavli IPMU (WPI), The University of Tokyo, 5-1-5 Kashiwanoha, Kashiwa 277-8583, Japan\\
$^{3}$Geneva Observatory, Geneva University, CH-1290 Sauverny, Switzerland\\
$^{4}$Department of Physics, Faculty of Science, Universiti Malaya, 50603 Kuala Lumpur, Malaysia\\
$^{5}$ Institut d'Astronomie et d'Astrophysique, Universit\'e Libre de Bruxelles (ULB), CP 226, B-1050 Brussels, Belgium
}

\date{Accepted XXX. Received YYY; in original form ZZZ}

\pubyear{\the\year{}}

\begin{document}
\label{firstpage}
\pagerange{\pageref{firstpage}--\pageref{lastpage}}
\maketitle

\begin{abstract}
The initial mass and metallicity of stars both have a strong impact on their fate. Stellar axial rotation also has a strong impact on the structure and evolution of massive stars. In this study, we exploit the large grid of GENEC models, covering initial masses from 9 to 500~$M_{\odot}$~ and metallicities ranging from $Z=10^{-5}$ (nearly zero) to 0.02 (supersolar), to determine the impact of rotation on their fate across cosmic times. Using the carbon-oxygen core mass and envelope composition as indicators of their fate, we predict stellar remnants, supernova engines, and spectroscopic supernova types for both rotating and non-rotating stars. We derive rates of the different supernova and remnant types considering two initial mass functions to help solve puzzles such as the absence of observed pair-instability supernovae.  We find that rotation significantly alters the remnant type and supernova engine, with rotating stars favouring black hole formation at lower initial masses than their non-rotating counterparts. Additionally, we confirm the expected strong metallicity dependence of the fates with a maximum black hole mass predicted to be below 50~$M_{\odot}$ at SMC or higher metallicities. A pair-instability mass gap is predicted between about 90 and 150~$M_{\odot}$, with the most massive black holes below the gap found at the lowest metallicities. Considering the fate of massive single stars has far-reaching consequences across many different fields within astrophysics, and understanding the impact of rotation and metallicity will improve our understanding of how massive stars end their lives, and their impact on the universe.
\end{abstract}

\begin{keywords}
stars: evolution; 
stars: massive; 
stars: rotation; 
stars: neutron; 
stars: black holes;
supernovae: general. 
\end{keywords}


\section{Introduction}

Understanding the fate of massive stars has far-reaching implications for stellar evolution, the formation of compact objects, and the classification of supernovae. As progenitors of neutron stars, black holes, and supernovae, these stars occupy a central position in the astrophysical landscape. Despite significant advancements in modelling stellar outcomes, key parameters such as rotation and metallicity remain under explored. These factors, however, exert significant influence over the evolutionary pathways and final fates of massive stars. The `fate' of a massive star is a description of the star during and after death, including the type of supernova explosion (if any) and the type of compact remnant (if any).

Metallicity and rotation affect key aspects of stellar evolution, including mass loss, angular momentum, and nucleosynthesis. Parameters such as the carbon-oxygen (CO) core mass and the hydrogen and helium composition of the envelope determine the type of stellar remnant and any supernovae produced. Previous research has addressed many aspects of stellar evolution, remnant types, and supernova classification, but the combined effects of mass, metallicity, and rotation require further investigation. 

Our primary aim is to determine the final fate of massive stars from the properties of stellar models at the end of core helium burning. The effects of initial mass, metallicity and rotation on the fate of massive stars are explored, and the role that different processes have on the final stages of evolution is considered. This is achieved using 1D stellar evolution models from the ongoing series of GENEC grids. These rotating and non-rotating models at metallicities from extremely metal poor (EMP), that of the Small Magellanic Cloud (SMC) and Large Magellanic Cloud (LMC) to solar and supersolar, ranging from 9 to $500$ \msol~ are analysed and their collective properties used to predict their final fate. Finally, statistical analysis of the above results is presented to provide an overview of the fate of massive stars at a stellar population level.

The structure of this paper is as follows: Section \ref{massive_star_models} introduces the physical ingredients of the massive star models and their properties, Section \ref{section:remnant} considers the different remnant types and BH mass distribution, Section \ref{section:sn} considers the different supernova types and Section \ref{section:pop} applies the results presented in Sections \ref{section:remnant} and \ref{section:sn} to a stellar population. This is followed by a discussion of wider implication of the work and conclusions in Sections \ref{discussion} and \ref{conclusion}.

\section{Massive star models}
\label{massive_star_models}
Complete and homogeneous grids of stellar models enable the analysis of a wide range of observations, and allow for the exploration of how stellar evolution depends on parameters such as initial mass, metallicity and rotation. They are also useful for considering the evolution of progenitors of neutron stars, supernovae and black holes, and also the evolution of galaxies. They are also a source of enrichment to the Universe in heavy elements. Many peculiarities of chemical abundances in galaxies find their origin in the different courses of stellar evolution, with different interplays between rotation, mixing and mass loss according to initial mass and metallicity. Here we concentrate on the differences in various evolutionary schemes. In this work, we use rotating and non-rotating stellar models at $\mathrm{Z} = 10^{-5}$ \citep{sibony2024}, 0.002 \citep{georgy2013},
0.006 \citep{eggenberger2022},
0.014 \citep{ekstrom2012}
and 0.02 \citep{yusof2022}
ranging from 9 to $500~M_{\odot}$~from the collection of GENEC grids, plus additional models for very massive stars \resub{(VMSs, $M_{\rm ini}>100\,M_\odot$)} \citep{yusof2013,martinet2023}, alongside models that have been calculated for this work that have not been published previously. A summary of the models and their origin is given in Table \ref{table:models} (in  Appendix \ref{appendix:models}). All of the models used ran until at least the end of core helium burning, and this is the stage of evolution at which the properties are calculated.

    \subsection{Physical ingredients of the models}
    The grids of models have been computed with the same input physics and physical ingredients to allow for direct comparison of their properties across masses and metallicities, facilitating determination of the fate of massive stars across cosmic time, with the exception of those from \citet{martinet2023} which have slightly different input physics for modelling VMSs. These ingredients are summarised below.

        The initial abundances of each grid of models and the mixture of heavy elements used in each grid is given by \citet{ekstrom2012}, with the absolute abundances scaled to the metallicity considered. The nuclear reaction rates are mainly taken from the NACRE database \citep{angulo1999}, and some have been updated as detailed in \citet{ekstrom2012}. Opacities are taken from OPAL \citep{iglesias1996}, used with low temperature opacities from \citet{ferguson2005} adapted for the high Ne abundance. 
        Convective zones are determined using the Schwarzschild criterion, and the convective core is extended with an overshoot parameter $d_{\mathrm{over}}/H_P = 0.1$ for convective H and He-burning cores. \resub{The non-rotating 180 and 300\,$M_\odot$  at $Z=0.006$ taken from \citet{martinet2023} instead use the Ledoux criterion and $d_{\mathrm{over}}/H_P = 0.2$ as well as an updated equation of state for the very late phase. Figure\,\ref{fig:m_co} shows that these two models fall in line with the surrounding models from the other sources and that the different input physics used in these models do not change the results presented in this work. Note nevertheless that} convection plays a pivotal role in both the evolution of massive stars and their fate, and so is one important source of uncertainty \resub{ \citep[see e.\,g.][and references therein]{kaiser2020}, which we will study in detail in Whitehead et\,al (in prep.)}. The rotating models in this work start on the ZAMS with $v_{\mathrm{ini}}/v_{\mathrm{crit}}=0.4$, \resub{where $\upsilon_\text{crit}=\sqrt{\frac{2}{3}\frac{GM}{R_\text{pb}}}$ and $R_\text{pb}$ is the polar radius at the critical limit}. This value is chosen as it aligns with the peak of the velocity distribution of young B stars in \citet{huang2010}. The treatment of rotation within the code is described in detail in \citet{maeder2012}. 
        
        A series of empirical and theoretical mass loss prescriptions are employed at different domains during the evolution to give the mass loss rate of the star. On the main sequence, the mass loss rate from \citet{vink2001} is used in the domains where it is valid and that from \citet{de1988} is used elsewhere. The formula from \citet{reimers1975} is used for RSG under $12$~\msol~and that from \citet{de1988} is used again for stars above $15$~\msol~with $\log{T_{\mathrm{eff}}}>3.7$. When $\log{T_{\mathrm{eff}}}\leq 3.7$, a fit of the data from \citet{van1999} and \citet{sylvester1998} is used \citep{crowther2000}. Mass loss rates for WR stars are given by \citet{nugis2000}, 
        \resub{ or the \citet{grafener2008} recipe in the small validity domain of this prescription. In some cases, the WR mass-loss rate of \citet{grafener2008} is lower than the rate of \citet{vink2001}. In these cases, we use the \citet{vink2001} prescription instead of that of \citet{grafener2008}.}
        Radiative mass loss has a metallicity dependence given by Eq. (\ref{z_ml}).
        \begin{align}
            \label{z_ml}
            \dot{M}(Z) = \dot{M}(Z_{\odot})(Z/Z_{\odot})^{\alpha}
        \end{align}
        where $\alpha = 0.85$ is used for the O-type phase and WN phase, $\alpha = 0.66$ for the WC and WO phases, and $\alpha=0.5$ is used for the \citet{de1988} prescription. There is no metallicity dependence for the mass loss rates of RSG stars (when $\log{T_{\mathrm{eff}}}\leq 3.7$). When considering WR stars, the initial metallicity is used in this equation rather than the surface metallicity \citep{eldridge2006}. A correction factor is applied to the mass loss rate of rotating models as per \citet{maeder2000}.

    \subsection{Properties of the models}
    \label{section:properties}
    \subsubsection{Surface properties}
\begin{figure*}
	\includegraphics[width=0.98\textwidth]{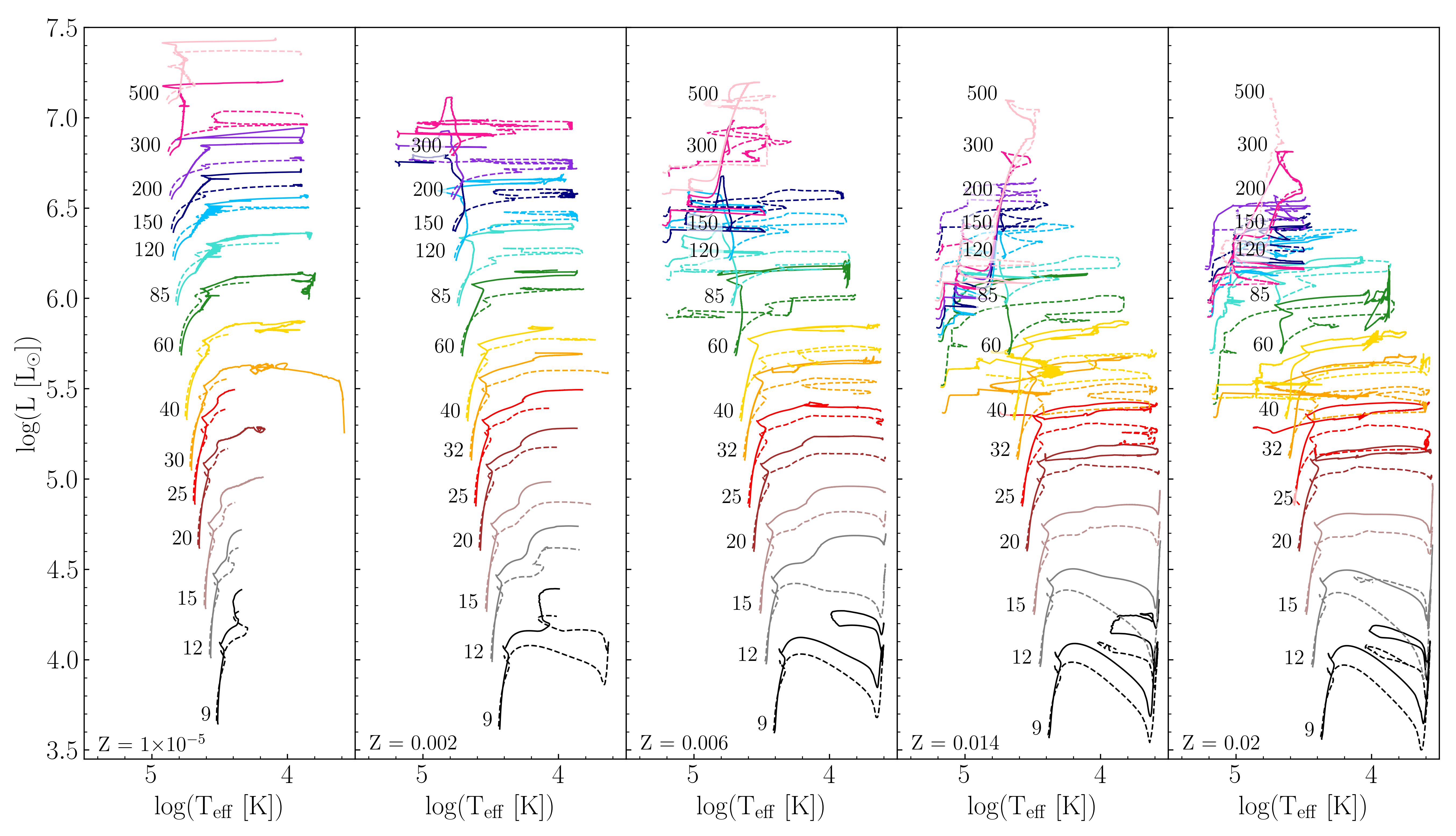}
        \caption{Evolutionary tracks in the HRD for rotating (solid line) and non-rotating (dashed line) models. The metallicity (increasing from left to right) of the different grids is given at the bottom of each panel. Each track is labelled and colour coded with its initial mass.}
        \label{fig:hrd}
    \end{figure*}

    The evolutionary tracks in the \resub{Hertzsprung-Russell diagram (HRD)} of all models used in this work are presented in Fig.~\ref{fig:hrd}, showing that the width of the main sequence band generally increases with the initial mass of the model at all metallicities. Rotating models are generally cooler and more luminous than non-rotating models with the same initial mass and metallicity. As the metallicity increases, massive star models at the same initial mass are generally cooler, due to metals at the surface increasing the opacity. \resub{The evolution of VMSs is dominated by mass loss and its metallicity dependence. At high metallicity, mass loss is large and the mass and luminosity of VMS models goes down, whereas at low metallicities, mass loss becomes gradually weaker. VMS models at very low metallicities reach very high luminosities and even expand late in their evolution.  } 
    The most luminous models are those with high initial mass and low metallicity that rotate. \resub{Such models may reach the Eddington limit and we will study this topic in more detail in Ismail et\,al (in prep.).}
    
    The hook found between the end of hydrogen burning and the start of helium burning is less noticeable for models with lower metallicity. This is because the temperature of the core at the end of hydrogen burning is higher in stars with lower metallicity, and so less contraction is required to heat up the core and maintain energy generation, resulting in a smaller hook feature for lower metallicities. Models at very high mass with $Z > 0.002$ evolve almost vertically in the HRD, as shown in Fig.~\ref{fig:hrd}, evolving across a wide range of luminosities but with almost constant effective temperature \citep[see also][]{yusof2013,higgins2022}. In these models, the convective core typically accounts for a large proportion of the mass of the star.

    The increased mixing in rotating models brings more hydrogen into the core, resulting in longer main sequence lifetimes at all initial masses and metallicities. This is not the case for the helium burning lifetime, as this tends to increase due to the convective core having a higher mass. More details about the models such as  tables of main sequence and helium burning lifetimes at different metallicities can be found in the corresponding GENEC grid papers \citep{ekstrom2012, georgy2013, yusof2013, yusof2022, martinet2023, sibony2024}.
 \subsubsection{Final mass}
    The final total mass is an indicator of the amount of mass lost throughout a star's life, and when coupled with the core and envelope masses it can be used to compare the impact of mass loss early and late in the evolution. We define the final mass as the total mass at the end of core helium burning, which is strongly dependent on the mass loss history of the model, and so it follows that it is dependent on the metallicity of the model. Taking the final mass at the end of helium burning may seem premature, as the star will continue to evolve and burn carbon, neon, oxygen and silicon before reaching a true final mass. However, these advanced phases of evolution have much shorter lifetimes and so any mass loss resulting from the usual mass loss recipe experienced after this point is very small and thus assumed to be negligible. Some mass loss may occur just before core collapse. This mass loss is not accounted for here as there is yet no physical description of it \citep{neilson2011}. As a result, the true final mass of these models might in some cases be smaller than the final mass taken at the end of helium burning, as quoted in this work. Rotation generally decreases the final mass when initial mass and metallicity are held constant, due to the increased mass loss rates experienced by rotating models. Similarly, due to the metallicity dependence of mass loss on the main-sequence, an increase in metallicity results in a decrease in final mass.
    
    Table \ref{table:data} in Appendix \ref{appendix} gives the final total mass, $M_{\mathrm{fin}}$, of all models used in this work. Both the rotating and non-rotating models with $Z=10^{-5}$ have the highest final mass across all initial masses, and non-rotating models with $Z=0.02$ have the lowest final mass across all initial masses. Rotating models generally have lower final masses due to the increased mass loss in rotating models. The significance of this effect increases with initial mass and metallicity until a peak at $Z=0.006$. \resub{At solar and supersolar metallicities, both non-rotating and rotating models experience significant mass loss, mainly post-MS below $\sim 50$~\msol but also more and more on the MS as the initial mass increases so the final mass and the final versus initial mass relation is no longer monotonic with respect to initial mass or metallicity}.
    
     \subsubsection{CO core mass}
\begin{figure}
	\includegraphics[width=0.49\textwidth]{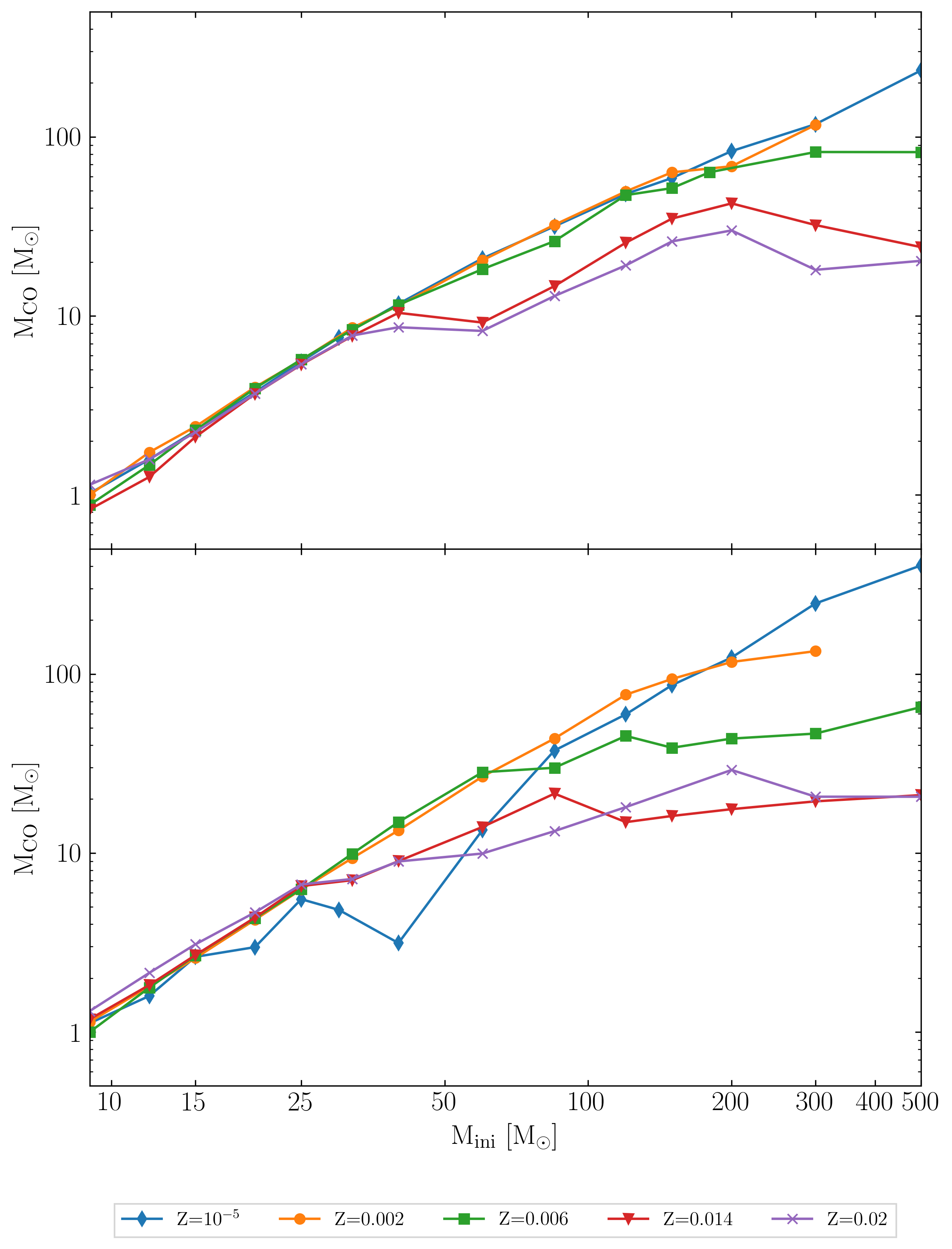}
        \caption{The CO core mass, $M_{\mathrm{CO}}$, of the non-rotating (top) and rotating (bottom) models at each metallicity, Z.}
        \label{fig:m_co}
    \end{figure}

    The carbon-oxygen (CO) core mass is one of the main indicators of fate used in this work. It is defined at the end of core helium burning (when the central helium mass fraction drops below $10^{-5}$)
    as the mass coordinate where the helium mass fraction falls below 1\% for the first time, when considering composition from the surface to the centre of the star. It corresponds to the maximum convective core mass reached by the end of core helium burning. This mass coordinate marks the edge of the core, where there is a steep density gradient which eventually helps the supernova shock-wave to eject material above the edge of the core. It is important to note that there are other ways to define the CO core mass, such as the mass coordinate where the combined carbon and oxygen mass fraction is greater than 75\% \citep{hirschi2004massive,hirschi2004}. This second definition finds a CO core mass that lies in-between the alpha core mass and the CO core mass as defined in this work, and it can include the helium burning shell outside of the core. It is for this reason that this definition is not used in this case, as it may suggest that the core is helium free when it is not. Fig. \ref{fig:m_co} shows how the CO core mass, $M_{\mathrm{CO}}$, varies with initial mass and metallicity.
    The CO core mass increases monotonically for non-rotating models with $M_{\mathrm{ini}}<30$~\msol, across all of the metallicities considered. Above this initial mass, models use a variety of mass loss prescriptions across different stages of their evolution and mass loss is much stronger, so the CO core mass still increases with initial mass, but the relationship is no longer monotonic.
    
    With the exception of models at $Z=10^{-5}$, the rotating models show a qualitatively similar relationship between CO core mass and initial mass as their non-rotating counterparts. At this metallicity, an unusual growth in the hydrogen burning shell causes a reduction in the CO core mass, as discussed in \citet{sibony2024}. This effect can be seen in rotating models with $Z=10^{-5}$ with initial mass $M_{\mathrm{ini}}=20-60~M_{\odot}$~to varying extents, and is most noticeable in the $30~M_{\odot}$~and $40~M_{\odot}$~models. They have significantly lower CO core masses than their non-rotating counterparts. For all $Z>10^{-5}$, rotating models generally have a higher CO core mass (due to rotation-induced mixing), and the dependence of CO core mass on metallicity is less clear than that of final mass.
     \subsubsection{Envelope mass}
    The mass and composition of the envelope is crucial in determining the type of supernova explosion that may occur after collapse. The mass of hydrogen and helium in the envelope at the end of core helium burning is given in Table \ref{table:data} in Appendix \ref{appendix}. Rotating models have a smaller hydrogen envelope mass for a given initial mass and they have a much smaller helium envelope mass across all metallicities. These relationships are much less monotonic than those for the final and CO core mass, suggesting that they are affected by a complex combination of many factors, including rotation, metallicity, mass loss and the extent of mixing. This has significant consequences when considering the supernova type of the models that are predicted to explode successfully, including as pulsation pair-instability supernova (PPISN) and pair-instability supernovae (PISN, see Sect.\,\ref{section:pisn}).
    
    \subsection{Compactness and the CO core mass}

    The advanced phases of the evolution of massive stars are largely determined by the CO core mass and the abundance of $^{12}\mathrm{C}$ at the end of core helium burning \citep{xin2025, chieffi2020, patton2020}. In particular, the CO core mass is significant in determining the further evolution of the star, but the abundance of $^{12}\mathrm{C}$ left after core helium burning is also informative as it determines the extent of both core and shell carbon burning phases. This mass fraction is not independent of the CO core mass \citep{chieffi2020}, and so will not be considered separately in this work. In order to relate the CO core mass to different remnant types, the compactness of the pre-supernova stellar core is often used; given by Eq.~(\ref{eqn:compactness}) evaluated at $M=2.5~M_{\odot}$~\citep{o2011}.  
    \begin{equation}
    \centering
        \label{eqn:compactness}
        \xi_M = \frac{M/M_{\odot}}{R(M)/1000\mathrm{km}}
    \end{equation}
    The compactness is a non-monotonic function of the CO core mass and is important when considering the final fate of massive stars. And so, the CO core mass at the end of core helium burning can be used to predict the type of compact remnant left behind when a massive star dies -  either a neutron star, black hole or no remnant in the case of PISN.
    
    Massive stars with $M_{\mathrm{CO}}<6~M_{\odot}$~~are thought to explode successfully and form neutron stars \citep{patton2020}. \citet{o2011} found that when $\xi_{2.5} > 0.45$, successful explosions are much less likely and there is a transition between neutron star and black hole formation; this will be referred to as the explodability limit. Similarly, \citet{ugliano2012} found that there is a transition region between neutron star and black hole formation when $0.15 < \xi_{2.5} < 0.35$. There is a peak in compactness between $6 < M_{\mathrm{CO}} < 12~M_{\odot}$~where the compactness increases and falls within this transition region, as per \citet{sukhbold2014}. When $6 < M_{\mathrm{CO}} < 8~M_{\odot}$~, the compactness increases through the transition region and above the explodability limit. Models with CO core mass within this range models are therefore considered as unlikely to explode and are expected to form black holes, with the possibility of a successful explosion to form a neutron star. Black holes formed within this transition mass range result from `failed' explosions and so form by fallback. 

    When $8 < M_{\mathrm{CO}} < 12~M_{\odot}$, the compactness falls below the explodability limit into the transition region and so such models are likely to explode successfully and are expected to form neutron stars, with the possibility of a failed explosion leading to black hole formation. This `island' of explodability is shown clearly in Fig. 13 of \citet{sukhbold2016} and is also eluded to in \citet{o2011}. It is important to note that later studies from \citet{wang2022, boccioli2023,maltsev2025} also find islands of explodability, but at very different masses and there is some dependence on metallicity. This highlights uncertainties in explodability predictions, in particular for high compactness models \citep{boccioli2024} but the CO mass still represent a reasonable indicator of the fate of massive stars. 
    
    When $12 < M_{\mathrm{CO}}<40~M_{\odot}$, a direct collapse to a black hole with $M_{\mathrm{BH}} = M_{\mathrm{fin}}$ is expected, meaning that there is no explosion at all \citep{patton2020}. Above $M_{\mathrm{CO}} = 40~M_{\odot}$, stars are expected to undergo PPISN followed by a core-collapse supernova, resulting in the formation of a black hole, and when $60 < M_{\mathrm{CO}} < 130~M_{\odot}$~stars are expected to be fully disrupted in a PISN that leaves behind no remnant \citep{farmer2019}. Hence, the CO core mass is also important in determining the supernova engine, as well as for the remnant type. Above $M_{\mathrm{CO}} = 130~M_{\odot}$, the photodisintegration instability, caused by the breaking apart of nuclei, allows for direct black hole formation again \citep{heger2003}. The values of $M_{\mathrm{CO}}$ corresponding to different remnant types are summarised in Table \ref{table:co}.
    
    \begin{table}
\caption{Dependency of remnant type on the CO core mass. PPISN refers to pulsation pair-instability supernovae and PISN refers to pair-instability supernovae.}
\centering
\label{table:co}
\begin{tabular}{cccccccc}
\hline
 &  Remnant type \\
\hline
$ M_{\mathrm{CO}} < 6 ~M_{\odot}$ & Neutron star \\
$ 6 < M_{\mathrm{CO}} < 8 ~M_{\odot}$ & Black hole (neutron star) \\
$ 8 < M_{\mathrm{CO}} < 12 ~M_{\odot}$  & Neutron star (black hole) \\
$ 12 < M_{\mathrm{CO}} < 40 ~M_{\odot}$ & Black hole \\
$ 40 < M_{\mathrm{CO}} < 60 ~M_{\odot}$ & PPISN with black hole \\
$ 60 < M_{\mathrm{CO}} < 130 ~M_{\odot}$ & PISN with no remnant\\
$ M_{\mathrm{CO}} > 130 ~M_{\odot}$ & Black hole \\
\hline
\end{tabular}
\end{table}

    \subsection{Envelope composition and spectroscopic supernova type}
\label{section:sn-types}
Supernova types are based on both the spectral and light curve properties of a supernova explosion. In this work, this is based on the composition of the envelope which is retained by the star (if the envelope has not been completely lost). Alternatively, the surface mass fraction of hydrogen/helium could also be used to determine supernova type, with \citet{yoshida2011} using $X_{\mathrm{He}}^{\mathrm{surf}} = 0.5$ as the boundary between Type Ib and Ic, but this measure is less widely used than envelope masses. It is largely agreed that the threshold amount of hydrogen when differentiating between Type II and Type Ib supernovae is low, such that~\citet{wellstein1999}, \citet{heger2003}, \citet{yusof2013} and \citet{yoon2010} use a threshold of $M_{\mathrm{H}}^{\mathrm{env}} < 0.5~M_{\odot}$~ to determine whether a star is free of hydrogen whereas \citet{georgy2009} use $M_{\mathrm{H}}^{\mathrm{env}} < 0.6~M_{\odot}$. Considering this, the choice $M_{\mathrm{H}}^{\mathrm{env}} < 0.5~M_{\odot}$~has been made in this work and so stars with $M_{\mathrm{H}}^{\mathrm{env}} < 0.5$~\msol~are considered `H-poor'. It is important to note that it is suggested in \citet{georgy2009} that a range of $0.6 < M_{\mathrm{H}}^{\mathrm{env}} < 1.5~M_{\odot}$ gives very similar results for supernova type. The threshold for hydrogen poor/rich (and so between Type II and Ib supernovae) is non-zero because the absence of H lines in spectra does not indicate a complete absence of hydrogen in the envelope; factors such as the temperature and density of the envelope are also important when considering the strength of the H lines \citep{dessart2012}. \resub{Type IIb supernovae are included as an intermediate type, for stars with  $0.033 <M_{\mathrm{H}}^{\mathrm{env}} < 0.5$~\msol~\citep{hachinger2012, gilkis2022}.}

It is more difficult to choose a threshold amount of helium to distinguish models which explode as Type Ic from Type Ib since it is thought that the absence of He lines in spectra may not indicate absence of helium in the envelope of the progenitor, since this helium may be hidden due to very low $^{56}$Ni mixing (if any) \citep{dessart2012}. Despite this, both \citet{frey2013} and \citet{liu2016} determine that progenitors of Type Ic supernovae are completely free of helium. \citet{frey2013} used a mixing algorithm based on 3D hydrodynamic simulations of massive stars to determine that the rates of mixing are higher than thought by \citet{dessart2012}. This mixing brings the helium into deeper, hotter layers of the star where it is burned to give O, resulting in completely helium free progenitors for Type Ic supernovae. In this work, stars with $M_{\mathrm{He}}^{\mathrm{env}} < 0.5~M_{\odot}$~are considered `He-poor', in alignment with the value chosen for $M_{\mathrm{H}}^{\mathrm{env}}$ (see Table\,\ref{table:sn}). It is important to note, however, that it is difficult to distinguish between Type Ib/c supernovae using the envelope composition alone, since recent studies indicate that a progenitor could have significant amounts of He, but it may not appear in the spectrum \citep{van2024}. 

To obtain the mass of hydrogen/helium in the envelope, the H/He mass fraction of each model was integrated throughout the star, since the CO core is free of both hydrogen and helium by definition. This allowed for determination of supernova type as described above. The derived values of $M_{\mathrm{H}}^{\mathrm{env}}$, $M_{\mathrm{He}}^{\mathrm{env}}$ and supernova type for each model are given in Table\,\ref{table:data}.
Note that models, which are predicted to directly collapse to black holes in Section \ref{section:remnant} are not allocated a type since they do not explode, while models resulting in BH (NS) or NS (BH) are allocated a type since it is uncertain to what extent they would explode (if at all). 
\begin{table}
\caption{Progenitor properties for different types of core-collapse SN. H/He envelope mass at the end of core helium burning is given by $M_{\mathrm{H/He}}^{\mathrm{env}}$ respectively.}
\centering
\label{table:sn}
\begin{tabular}{cccccc}
\hline
 \multicolumn{2}{c}{Envelope composition} &  SN type \\
\hline
$M_{\mathrm{H}}^{\mathrm{env}} > 2 ~M_{\odot}$ & $ M_{\mathrm{He}}^{\mathrm{env}} > 0.5 ~M_{\odot}$ & Type IIP \\
$ 0.5 < M_{\mathrm{H}}^{\mathrm{env}} < 2 ~M_{\odot}$ & $ M_{\mathrm{He}}^{\mathrm{env}} > 0.5 ~M_{\odot}$ & Type IIL \\
\resub{$ 0.033 < M_{\mathrm{H}}^{\mathrm{env}} < 0.5 ~M_{\odot}$} & \resub{$ M_{\mathrm{He}}^{\mathrm{env}} > 0.5 ~M_{\odot}$} & \resub{Type IIb} \\
\resub{$M_{\mathrm{H}}^{\mathrm{env}} < 0.033 ~M_{\odot}$} & $ M_{\mathrm{He}}^{\mathrm{env}} > 0.5 ~M_{\odot}$ & Type Ib \\
$M_{\mathrm{H}}^{\mathrm{env}} < 0.5 ~M_{\odot}$ & $ M_{\mathrm{He}}^{\mathrm{env}}< 0.5 ~M_{\odot}$ & Type Ic \\
\hline
\end{tabular}
\end{table}

\begin{figure*}
	\includegraphics[width=0.85\textwidth]{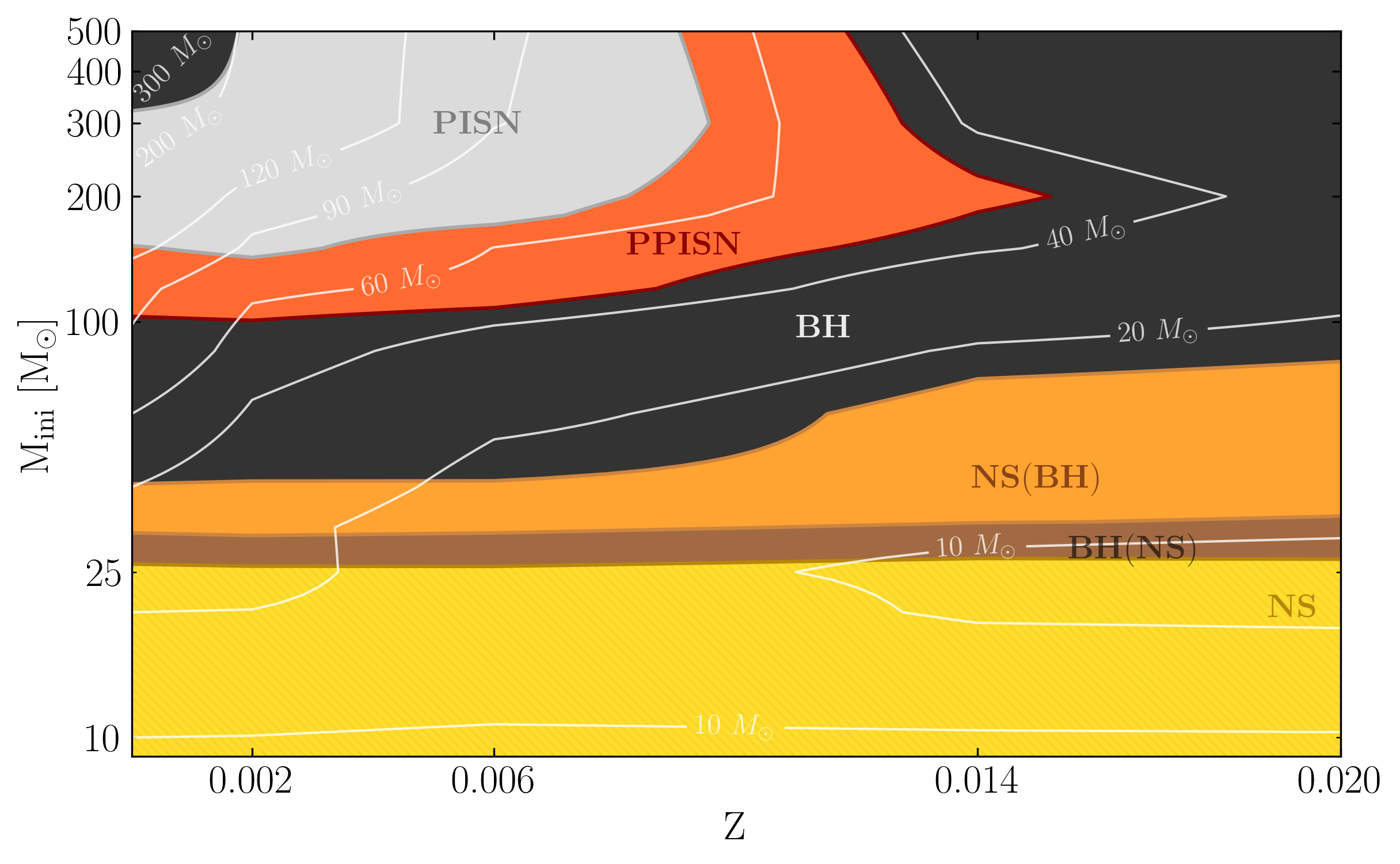}
    \caption{Remnant type of non-rotating massive star models as a function of initial mass and metallicity. The boundaries for each remnant type are given in Table \ref{table:co}, and the white contour lines indicate final mass.}
    \label{fig:rem_s0}
\end{figure*}
\begin{figure*}
	\includegraphics[width=0.85\textwidth]{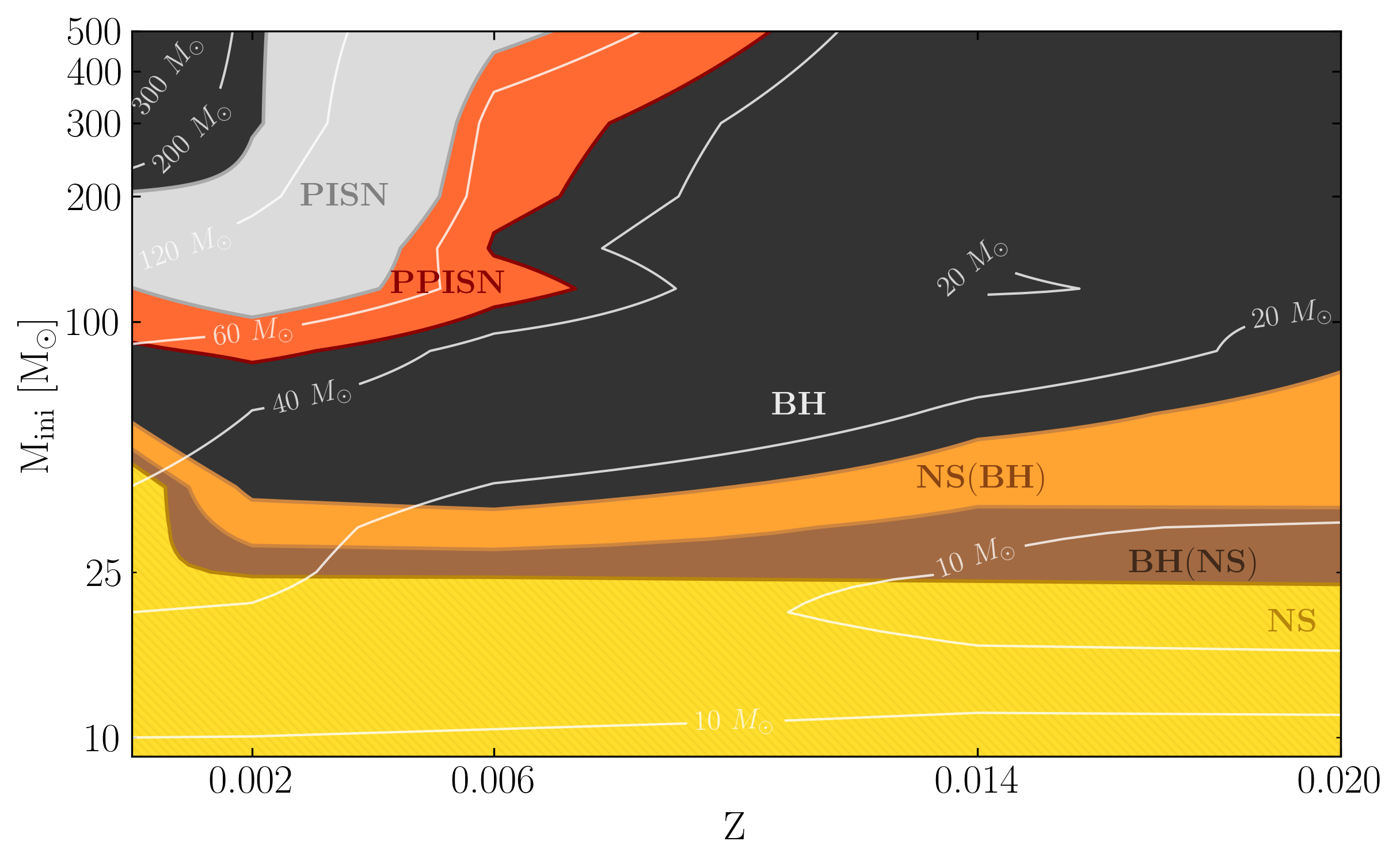}
    \caption{Same as Fig.\,\ref{fig:rem_s0} for rotating models.}
    \label{fig:rem_s4}
\end{figure*}

\section{Remnant type and BH mass distribution}
\label{section:remnant}
Contour plots exploring how the predicted remnant type varies with initial mass, metallicity and rotation are presented in this section, alongside a consideration of how black hole masses below the pair-instability gap are distributed. 

The CO core mass was calculated from each model in the ongoing series of grids, and linear interpolation between these values across an evenly spaced grid of initial masses (with steps of $1~M_{\odot}$) has allowed for analysis of the CO core mass across the whole mass range considered. This interpolation was also performed on the final mass and H and He envelope mass of the models. The use of linear interpolation may also result in missing features in the map. This is why it is also useful to refer back to the model data, given in Table \ref{table:data}, around interesting features in the contour map.

The rotating model with $M_{\mathrm{ini}}=500$~\msol, $Z=0.02$ did not reach the end of hydrogen burning due to numerical issues. Data at this point is essential for the interpolation function used, and it was assumed that the properties of this model would be the same as those of the rotating model at the same metallicity with $M_{\mathrm{ini}}=300$~\msol. This is a reasonable assumption to make, as the properties of the rotating~$300$~\msol~and $500$~\msol~models at $Z=0.014$ also converge to a very similar value.
    
The contour boundaries used in Figs. \ref{fig:rem_s0} and \ref{fig:rem_s4} are given in Table \ref{table:co}, and it is important to note that the behaviour of the variables may be less accurately represented at such boundaries, and that further analysis of these regions may be required to build an accurate picture of how the response variable is related to the independent variables.
    
\subsection{Effect of metallicity}
Firstly, the effect of changing initial mass and metallicity on the CO core mass, and so the remnant type, will be considered for non-rotating models. It is expected that stars with higher initial mass will generally have a higher CO core mass, if the metallicity is constant. It is expected that increasing the metallicity will lead to an increased rate of mass loss, due to the dependence given by Eq. (\ref{z_ml}). If this mass loss occurs early in the evolution, it will result in smaller helium core masses and so leading to a smaller CO core mass.

Fig. \ref{fig:rem_s0} shows the dependence of fate on initial mass and metallicity for non-rotating models. When initial mass is low, the dependence of remnant type on metallicity is very limited. This is due to very low rates of mass loss, particularly early in the evolution. Hence, the CO core mass for a particular initial mass is constant across the range of metallicities considered in this work. Massive stars with $M_{\mathrm{ini}}<25~M_{\odot}$ are predicted to end their lives as neutron stars (NS), for the whole metallicity range considered. At higher initial mass, when $25<M_{\mathrm{ini}}<30~M_{\odot}$, black holes via failed supernovae are expected, with the possibility of a successful explosion leading to a NS remnant, referred to as BH (NS), as per Section \ref{section:remnant}. When $30<M_{\mathrm{ini}}<40~M_{\odot}$, NS are expected but there is still the possibility of a black hole remnant, referred to as NS (BH). At higher initial mass, when $M_{\mathrm{ini}}>40~M_{\odot}$, the remnant type shows a strong dependence on metallicity.

At low metallicity and high initial mass, the CO core mass remains high due to low rates of mass loss early in the evolution. Then, for $40<M_{\mathrm{ini}}<100~M_{\odot}$, direct black holes are formed. As shown by the white contours on Fig.~\ref{fig:rem_s0}, the maximum black hole mass below the pair-instability gap ranges from $M_{\mathrm{BH}} \approx~30~-~90~M_{\odot}$ and so is highly dependent on metallicity \resub{(the BH mass distribution will be discussed in more detail in Section\,\ref{section:bh})}. The upper boundary for black hole formation without encountering the pair-instability \resub{(boundary between the black and orange regions)} increases with metallicity and so has an upward slope in  Fig. \ref{fig:rem_s0}. Likewise, the upper boundary for PPISN forming black holes below the PISN gap \resub{(boundary between the orange and grey regions)} is also dependent on metallicity, such that it is located at $M_{\mathrm{ini}}=150~M_{\odot}$~when $Z=10^{-5}$, and $M_{\mathrm{ini}}=300~M_{\odot}$~when $Z=0.01$. When $Z<0.002$, PISN are predicted from this boundary until $M_{\mathrm{ini}}=325~M_{\odot}$~above which direct black hole formation is predicted once again. When $Z>0.002$, PISN are predicted up to $M_{\mathrm{ini}}=500~M_{\odot}$, which is the highest mass considered in this work. 

At high metallicity and high initial mass, the CO core mass decreases as metallicity increases (assuming constant initial mass). This is because mass loss early in the evolution becomes significant, as these models evolve at higher luminosities than their lower mass counterparts; this effect is scaled with metallicity as per the metallicity dependence of mass loss. The boundary between NS (BH) and direct black hole formation lies between $50<M_{\mathrm{ini}}<80~M_{\odot}$ depending on the metallicity. Above this, models result in direct collapse to black holes, and when $Z>0.014$ this is the case for all models in this region. Finally, PPISN are predicted at $Z<0.014$ in the region of $M_{\mathrm{ini}}=200~M_{\odot}$, as shown in Fig. \ref{fig:rem_s0}.

\subsection{Effect of rotation}
The impact of rotation is complex as it has competing effects on the evolution. Increased mixing leads to the formation of larger helium cores, and so larger CO core masses are expected. On the other hand, rotation also leads to higher luminosities and so higher rates of early mass loss \resub{ as well as modest mechanical mass loss when stars reach critical rotation at low metallicities \citep[see][Sect.\,3.1 for more details]{sibony2024}.} This would lead to smaller helium cores and so a decrease in CO core mass. And so, there are two competing effects that both result from including rotation in the models. The increase in CO core mass due to internal mixing tends to be the dominant effect at lower initial mass and metallicity, whereas the decrease due to increased mass loss tends to dominate at higher initial mass and metallicity. Hence, the results are expected to show an interesting combination of these effects.

The impact of rotation can be seen by comparing Figs. \ref{fig:rem_s0} and \ref{fig:rem_s4}. The results on the $M_{\mathrm{ini}}-Z$ plane will be separated into four cases depending on the initial mass and metallicity, where the dominant effect due to rotation differs.

When $M_{\mathrm{ini}}<60~M_{\odot}$ and $Z<0.002$, Fig. \ref{fig:rem_s4} differs significantly from the same region in Fig. \ref{fig:rem_s0}. Firstly, unusual growth in the hydrogen burning shell, as discussed in \citet{sibony2024}, causes a decrease in CO core mass at $Z=10^{-5}$, up until $Z=0.002$. This effect is strongest around $40~M_{\odot}$, but can be seen in many models at $Z=10^{-5}$. This causes the location of the boundary for NS, BH (NS) and NS (BH) to increase in initial mass when compared to the non-rotating case, where this effect is not seen, introducing a negative slope to the boundary \resub{(the limiting mass between NS and BH(NS), for example, decreases when the metallicity increases)}. Hence, more stars are predicted to form NS, BH (NS) and NS (BH) than in the non-rotating case in this region. This is \textit{not} due to increased rates of mass loss (as one might expect), but is because of the interesting effect of rotation on the hydrogen burning shell, which is as discussed above. The final mass contours (white lines in Figs. \ref{fig:rem_s0} and \ref{fig:rem_s4}) are very similar as increased mass loss is not the dominant effect of \resub{rotation} in this region, and so both rotating and non-rotating models experience similar rates of mass loss.

When $M_{\mathrm{ini}}<60~M_{\odot}$ and $Z>0.002$, the dominant effect of rotation is increased mixing, leading to higher CO core masses than in the non-rotating case. This causes the boundary between NS (BH) and direct black hole formation to generally decrease in initial mass when compared to the non-rotating case. Similarly to the non-rotating case, the boundary increases to higher initial mass as metallicity increases due to the fact that CO core mass generally decreases with increasing metallicity in this region of Fig. \ref{fig:rem_s4}. In the rotating case, this increase is seen as a gentle upward slope in Fig. \ref{fig:rem_s4}, unlike the sharp increase and plateau seen in Fig. \ref{fig:rem_s0}. Again, the final mass contours are very similar between Fig. \ref{fig:rem_s0} and \ref{fig:rem_s4} in this region, since the rates of mass loss are not greatly impacted by rotation at low initial mass.

When $M_{\mathrm{ini}}>60~M_{\odot}$ and $Z<0.002$, the dominant effect of rotation is also increased mixing, leading to higher CO core masses when compared to results from Fig. \ref{fig:rem_s0}. When $Z=10^{-5}$, direct black holes are formed for $60 < M_{\mathrm{ini}} < 90~M_{\odot}$. The lower boundary of this mass range is significantly higher than that in the non-rotating case ($M_{\mathrm{ini}}=40~ M_{\odot}$) due to the interaction between the hydrogen burning shell and helium burning core. The upper boundary is lower in the rotating case due to an increase in CO core mass due to internal mixing.
    
The maximum black hole mass below the pair-instability gap ranges from $M_{\mathrm{BH}} \approx 35-60~M_{\odot}$, which is discussed further in Section \ref{section:bh}. The lower boundary of this mass range is lower than in the non-rotating case due to the hydrogen burning shell effect, and the upper boundary is significantly lower than in the non-rotating case due to increased mixing.

When $M_{\mathrm{ini}}>60~M_{\odot}$ and $Z>0.002$, the dominant effect of rotation is increased mass loss. The metallicity boundaries for PPISN and PISN are lower for rotating models and when $Z>0.01$, only direct black holes and NS (BH) are predicted in this region. When compared to the same region in Fig. \ref{fig:rem_s0}, the impact of rotation can be seen by the decreased final masses, which leads to lower black hole masses.

\subsection{The BH mass distribution}
\label{section:bh}
The remnant mass in the case of black hole formation, 
$M_{\mathrm{BH}}$, depends on the amount of mass assumed to be ejected during or following the final collapse.  
For stars undergoing PPISN (for models with $ 40 < M_{\mathrm{CO}} < 60 ~M_{\odot}$ in this work), there will be significant mass ejected due to the pulsations and so determining the black hole mass is subject to uncertainties. In this work, we consider that the black hole mass depends on the CO core mass and metallicity according to Eq. (\ref{eqn:bh}), which is adapted from \citet{farmer2019}.

    \begin{equation}
        \label{eqn:bh}
        M_{\mathrm{BH}}=a_1 M^2_{\mathrm{CO}} + a_2 M_{\mathrm{CO}} +a_3 \log{Z} + a_4
  \end{equation}
where $a_1 =-0.096$, $a_2 =8.564$, $a_3 =-2.07$, $a_4 =-152.97$.

Note that Eq. (\ref{eqn:bh}) is not used in this work for black hole masses of stars that undergo failed supernova and direct collapse as it is based on He stars only, and so~$M_{\mathrm{BH}}=M_{\mathrm{fin}}$ alone will be used for black holes formed by direct collapse below the PI mass gap ($12 < M_{\mathrm{CO}}<40~M_{\odot}$). Similarly, for black holes formed by direct collapse above the PI mass gap ($ M_{\mathrm{CO}} > 130 ~M_{\odot}$), ~$M_{\mathrm{BH}}=M_{\mathrm{fin}}$ is used. \resub{Note that mass ejection might be possible if a rotating star above the PISN mass gap forms an accretion disk around the BH that can drive jets \citep[see e.\,g.][for hydrodynamical models of such a scenario in PopIII VMSs]{ohkubo2006}. 
Such a mass ejection would
produce a BH of $M_{\mathrm{BH}} < M_{\mathrm{fin}}$, possibly reducing the upper limit of the mass gap
below 130 M$_\odot$, although we do not consider such mass ejection in the
present study}.

For stars undergoing a full PISN ($60 < M_{\mathrm{CO}} < 130~M_{\odot}$), the entire star is ejected and no remnant is left behind.
It is more complicated when considering failed supernova explosions, as some (or all) of the envelope could be ejected. In this work, $M_{\mathrm{BH}}=M_{\mathrm{CO}}$ is used for models that undergo failed explosions, with both NS (/BH; models with $ 8 < M_{\mathrm{CO}} < 12 ~M_{\odot}$) and BH (/NS; models with $ 6 < M_{\mathrm{CO}} < 8 ~M_{\odot}$) remnant types, assuming that they form black holes and eject their hydrogen and helium rich envelope. The relations used to determine the remnant mass in the case of black hole formation, 
$M_{\mathrm{BH}}$, are summarised in Table\,\ref{table:bh} and the BH masses obtained using these are listed in Table\,\ref{table:data} (column 5) for the relevant models.

    \begin{table}
\caption{Relations used to determine the remnant mass in the case of black hole formation, 
$M_{\mathrm{BH}}$.}
\centering
\label{table:bh}
\begin{tabular}{cccccccc}
\hline
Relevant $ M_{\mathrm{CO}}$ range &  $M_{\mathrm{BH}}$ \\
\hline
$ M_{\mathrm{CO}} < 6 ~M_{\odot}$ & no BH \\
$ 6 < M_{\mathrm{CO}} < 8 ~M_{\odot}$ & $ M_{\mathrm{CO}}$ \\
$ 8 < M_{\mathrm{CO}} < 12 ~M_{\odot}$  & $ M_{\mathrm{CO}}$ \\
$ 12 < M_{\mathrm{CO}} < 40 ~M_{\odot}$ & $M_{\mathrm{fin}}$ \\
$ 40 < M_{\mathrm{CO}} < 60 ~M_{\odot}$ & Eq.\,\ref{eqn:bh} \\
$ 60 < M_{\mathrm{CO}} < 130 ~M_{\odot}$ & no BH\\
$ M_{\mathrm{CO}} > 130 ~M_{\odot}$ & $M_{\mathrm{fin}}$ \\
\hline
\end{tabular}
\end{table}

\begin{figure}
	\includegraphics[width=0.5\textwidth]{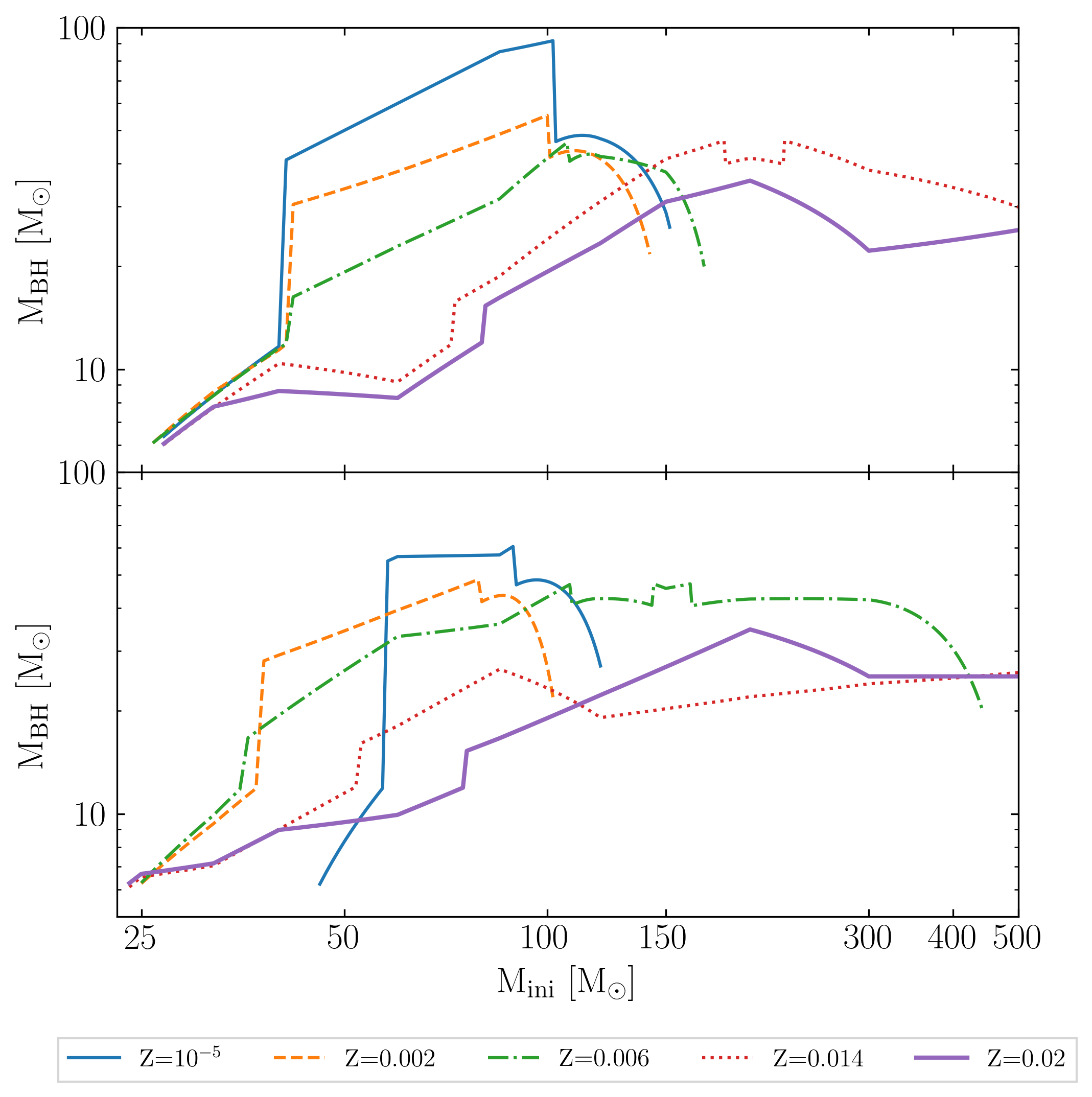}
    \caption{Black hole mass distribution for non-rotating (top) and rotating (bottom) models below the pair-instability mass gap.}
    \label{fig:bh_m}
\end{figure}

Table \ref{table:m_bh} gives the maximum black hole mass below the pair-instability gap  calculated using the values of $M_{\mathrm{fin}}$ and $M_{\mathrm{CO}}$ directly from the models, while Fig. \ref{fig:bh_m} shows the black hole mass distribution across different metallicities, calculated using the interpolated values of $M_{\mathrm{fin}}$ and $M_{\mathrm{CO}}$.

\resub{ The detection of the GW190521 gravitational wave event involving a black hole with mass~$85~M_{\odot}$ \citep{abbott2020} and GW231123 with a BH mass around $100~M_{\odot}$ \citep{gw231123} challenge the existence of a PI mass gap between $M_{\mathrm{BH}} \approx 50-130~M_{\odot}$~\citep{woosley2019,farmer2019}, but the results we present suggest a reduced mass gap with higher boundaries than this with a mass gap predicted between about 90 and 150~$M_{\odot}$. This was also found in \citet{vink2021}, where it is predicted that black holes of mass $\sim~90~M_{\odot}$ can form in low metallicity environments \citep[see also][]{farrell2021}. The lower end of the PISN mass gap presented in this work is also consistent with results from \citet{winch2024}, where a maximum black hole mass below the PI mass gap of $93.3~M_{\odot}$ was found using rotating models with $Z=10^{-3}$ as well as the lower end of 100~$M_{\odot}$ found by \citet{costa2025} for non-rotating models at very low metallicities. The highest BH masses below the gap come from models with a massive hydrogen-rich envelope and at the same time a CO core mass below the PPISN limit. Mass ejection due to pulsations and the CO core mass corresponding to the lower end of the PPISN mass range are two key uncertainties affecting predictions of the mass gap.}

It is important to also consider the mass of black holes \textit{above} the PI mass gap. As per Figs. \ref{fig:rem_s0} and \ref{fig:rem_s4}, direct black holes are predicted for both rotating and non-rotating models at low metallicities (below that of the SMC). These black holes are predicted to have~$M_{\mathrm{BH}}=M_{\mathrm{fin}}$, which is very close to their initial mass as the models lose little mass throughout their evolution. In the present grid, the least massive BH above the gap is 150~$M_{\odot}$ for the rotating 300~$M_{\odot}$ at $Z=0.002$ and the heaviest black hole predicted  has $M_{\mathrm{BH}}= 465.8~M_{\odot}$. It lies above the PI mass gap and originates from the non-rotating progenitor model at $Z=10^{-5}$ with $M_{\mathrm{ini}} = 500~M_{\odot}$. At the lower end of the mass range considered, no black holes below $6~M_{\odot}$ are predicted by design. Note that this is expected as there is another likely black hole mass gap observed at $2-5~M_{\odot}$ \citep{bailyn1998, farr2011, jonker2021}.

     \begin{table}
\caption{The maximum black hole mass below the pair-instability gap per metallicity, calculated using the values of $M_{\mathrm{fin}}$ and $M_{\mathrm{CO}}$ directly from the models. }
\centering
\label{table:m_bh}
\begin{tabular}{cccccccc}
\hline
 Z &  $v_{\mathrm{ini}}/v_{\mathrm{crit}}$ & $M_{\mathrm{BH, max}}$\\
\hline
$10^{-5}$ & 0   & 84.99 \\
 & 0.4 & 57.22 \\
0.002     & 0   & 48.75 \\
     & 0.4 & 43.51 \\
0.006     & 0   & 41.94 \\
     & 0.4 & 45.67 \\
0.014     & 0   & 41.44 \\
    & 0.4 & 26.49 \\
0.02      & 0   & 35.65 \\
      & 0.4 & 34.64 \\
\hline
\end{tabular}
\end{table}

\section{Supernova Type}
\label{section:sn}
    Contour plots exploring how the predicted supernova type varies with initial mass, metallicity and rotation are presented in this section, alongside a consideration of the main effects of metallicity and rotation.
    \subsection{Impact of metallicity}
 Firstly, the effect of changing initial mass and metallicity will be considered for non-rotating models. It is expected that stars with higher initial mass will generally lose more of their envelope as they evolve to higher luminosities, increasing the rate of mass loss which they experience. Additionally, stars with higher metallicity will lose more of their envelopes due to increased levels of mass loss, due to the dependence given by Eq. (\ref{z_ml}). Due to this, more Type II supernovae are expected at low metallicity, and more Type 1b/c supernovae are expected at higher metallicity.
\begin{figure}
	\includegraphics[width=0.5\textwidth]{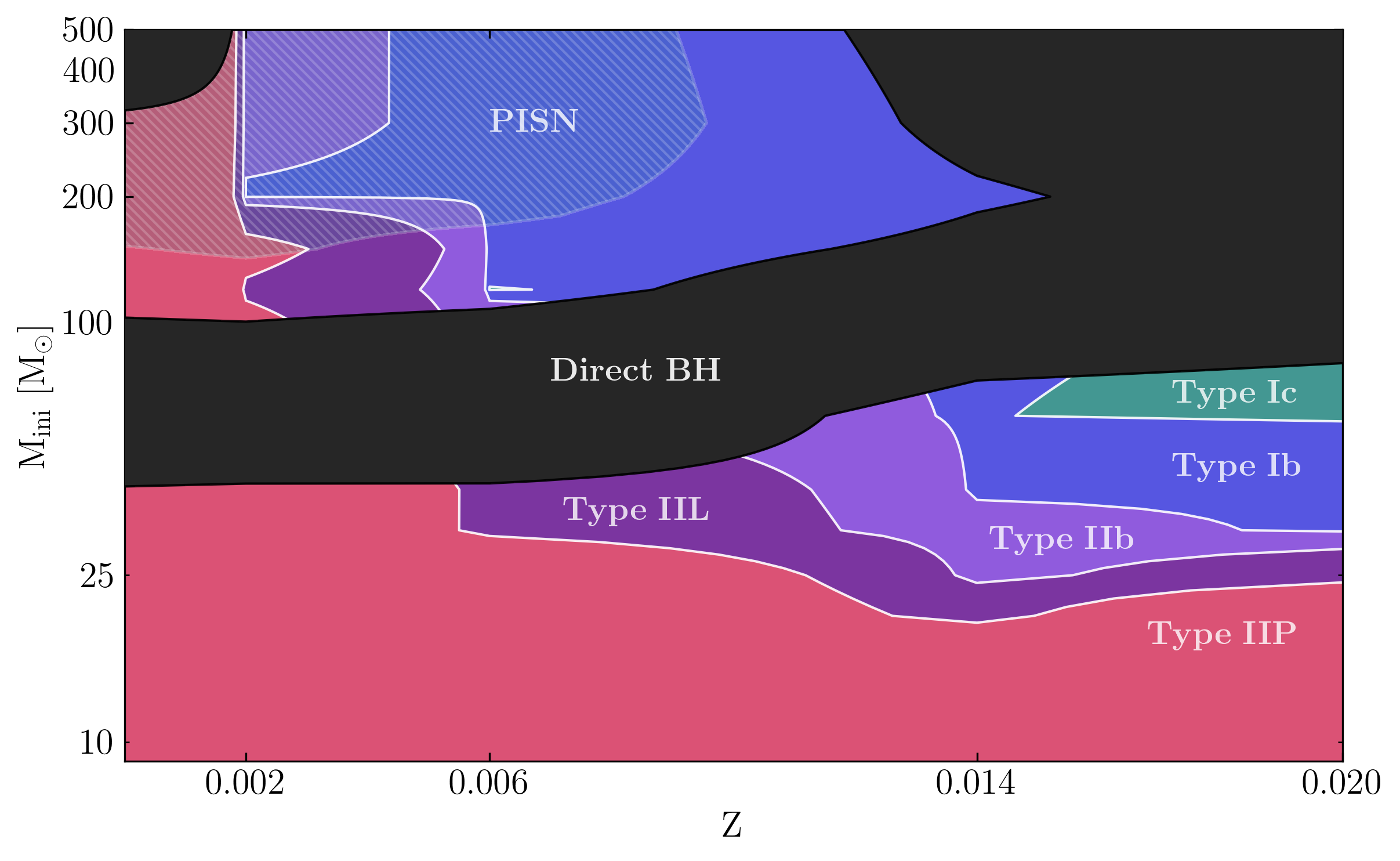}
    \caption{\resub{Supernova type of non-rotating massive stars as a function of initial mass and metallicity. The conditions used in this work to associate a given pre-supernova structure with a SN type is given in Table \ref{table:sn}.}}
    \label{fig:sn_s0}
\end{figure}
\begin{figure}
	\includegraphics[width=0.5\textwidth]{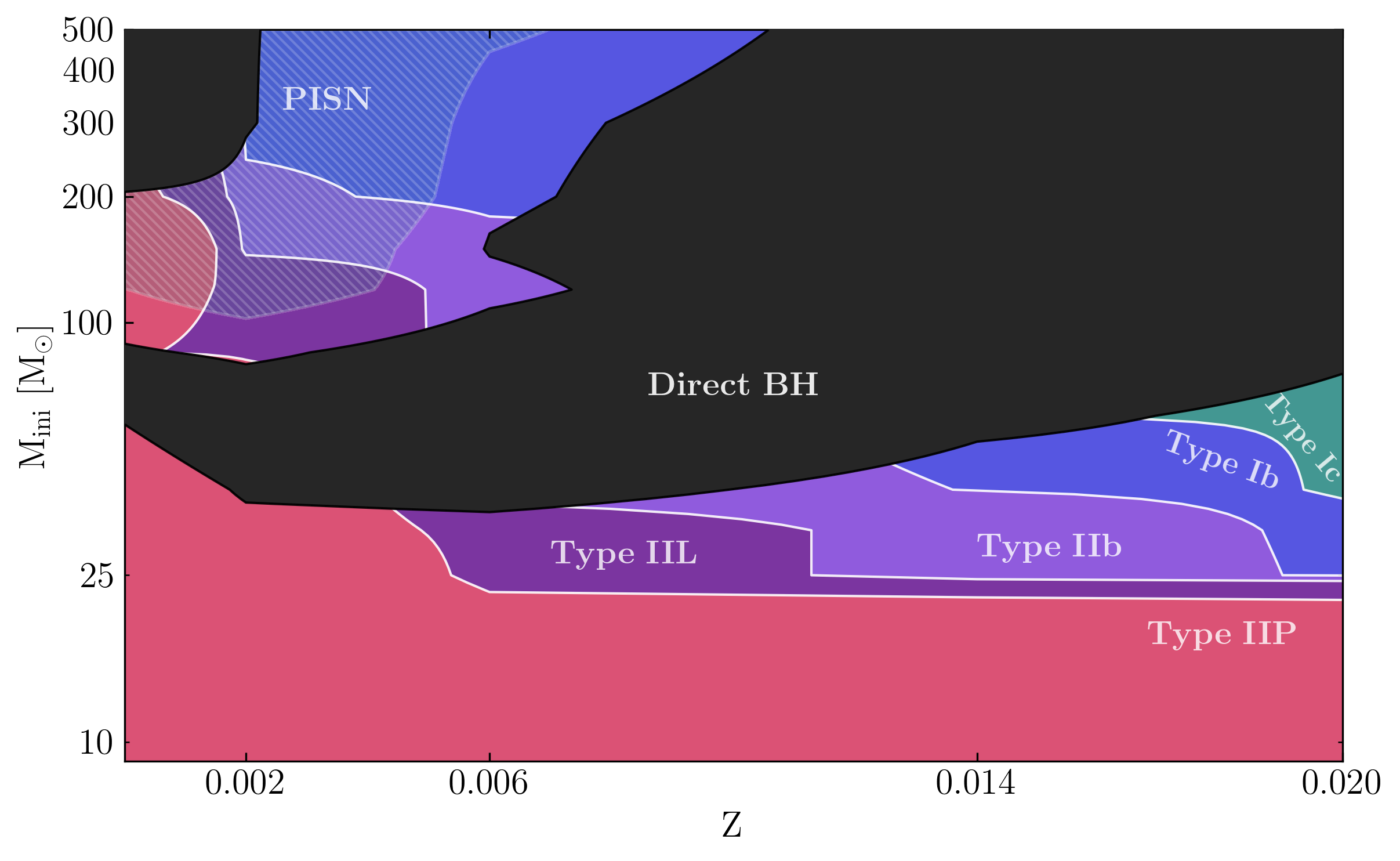}
    \caption{\resub{Same as Fig.\,\ref{fig:sn_s0} for the rotating models.}}
    \label{fig:sn_s4}
\end{figure}

 Fig. \ref{fig:sn_s0} shows that when initial mass is low ($M_{\mathrm{ini}}<20$~\msol), all massive stars are expected to explode as Type IIP supernovae. This is because they experience low levels of mass loss during their evolution and so retain most of their envelope, which is rich in both hydrogen and helium. Similarly, most massive stars with low metallicity ($Z<0.01$) are also expected to explode as Type IIP supernovae. This is due to the metallicity dependence of mass loss, meaning that these stars experience low levels of mass loss. Those that do not explode as Type IIP supernovae are expected to directly collapse to a BH. At the boundary between Type IIP and Ib supernovae in Fig. \ref{fig:sn_s0}, there is a region where stars retain a small amount of hydrogen in their envelope and so they are expected to explode as Type IIL \resub{ or IIb} supernovae. At higher initial mass, the dependence of supernova type on metallicity is stronger. When $M_{\mathrm{ini}}>40$~\msol~and $Z<0.01$, stars are predicted to either collapse directly to black holes, or explode as either PPISN or PISN of different types. PISN are indicated in Fig. \ref{fig:sn_s0} by the hatched region, then PPISN are expected to occur in the region between direct BH and PISN. 

 Type Ib supernovae are expected from $Z=0.002$ and above, depending on initial mass. As metallicity increases, we expect stars with lower initial mass to explode as Type Ib supernovae. This is because stars at lower initial mass tend to lose more of their envelope at higher metallicity. At higher metallicity, when $Z\geq 0.014$, initial mass has a more significant impact on supernova type than metallicity. This is shown in Fig. \ref{fig:sn_s0}, as the boundaries between different supernova types are largely horizontal, with a gentle upwards slope. This means that stars at supersolar metallicity explode as Type IIP supernovae at slightly higher initial mass than those at solar metallicity. \resub{Mass loss occurring earlier (faster) at supersolar metallicity means that cores are slightly smaller and that it is slightly harder to lose the envelope.} 
 
 When $50 < M_{\mathrm{ini}}< 70$~\msol, Type Ic supernovae are predicted above solar metallicity. These stars \resub{are free of hydrogen and (almost) free of helium}, since they experienced very high levels of post-MS mass loss. Stars between solar and supersolar metallicity above this mass range are expected to directly collapse to black holes. Hence, the impact of metallicity on supernova type is due to the dependence of mass loss rates on the metallicity. This effect changes depending on initial mass, which is related to the luminosity. 
\subsection{Impact of rotation}

Rotation is expected to decrease the minimum initial mass of progenitors of Type Ib/c supernovae as they will evolve to higher luminosities than their non-rotating counterparts. Similarly, rotation is expected to decrease the number of Type II supernovae predicted as more models will lose their envelopes due to increased levels of mass loss. 
The impact of rotation can be seen by comparing Figs. \ref{fig:sn_s0} and \ref{fig:sn_s4}, and differs depending on both initial mass and metallicity. This is due to the competing effects that rotation has on the mass of hydrogen and helium in the envelope. The results on the $M_{\mathrm{ini}}-Z$ plane will be separated into four cases depending on the initial mass and metallicity, where the dominant effect due to rotation differs. 

     When initial mass and metallicity are low, ($M_{\mathrm{ini}}<40$~\msol~and $Z<0.01$), the impact of rotation on supernova type is limited. The impact of rotation in this region of Fig. \ref{fig:sn_s4} is limited to $Z<10^{-5}$, due to the interaction between the hydrogen burning shell and helium burning core. The lower CO core mass means that models at lower initial mass that would directly collapse to black holes in the non-rotating case are predicted to successfully explode in the rotating case. 
     
     When initial mass is low and metallicity is higher ($M_{\mathrm{ini}}<40$~\msol~and $Z>0.01$),~rotation has a few interesting effects. Firstly, the boundaries between Type IIP, IIL and \resub{IIb} supernovae become completely horizontal, showing no dependence on metallicity. This is because models with $M_{\mathrm{ini}}\geq25$~\msol~completely lose their hydrogen envelope at both solar and supersolar metallicity. In the non-rotating case, the boundaries show more of a metallicity dependence, since models at solar and supersolar metallicity lose their hydrogen envelope at different initial mass. 
     
     On the other hand, the boundary between Type Ib and Ic is \textit{more} dependent on metallicity such that stars at supersolar metallicity explode as Type Ic supernovae at lower initial mass than their non-rotating counterparts. This is because more stars at supersolar metallicity lose both their hydrogen and helium envelope due to increased mass loss in the rotating case. 
     
     When initial mass is high and metallicity is low ($M_{\mathrm{ini}}>40$~\msol~and $Z<0.01$), \resub{the boundaries IIP/IIL/IIb/Ib move to lower metallicities in the rotating case.} This is because rotating stars evolve at higher luminosities and so experience higher rates of mass loss which can partially strip the star of its hydrogen envelope, even at metallicites below $Z=0.002$. Above $Z=0.006$, stars are predicted to directly collapse to black holes from $100$~\msol~to~$300$~\msol, whereas no direct black holes are predicted above $M_{\mathrm{ini}} = 100$~\msol~and $Z=0.006$ in the non-rotating case. Additionally, no Type Ic supernova are predicted in this region, whereas in the non-rotating case there is a single point where they are expected. 
     
     When initial mass and metallicity are both high ($M_{\mathrm{ini}}>40$~\msol~and $Z>0.01$), only a direct collapse to BH is predicted in the rotating case. This was not the case for non-rotating stars, where some stars in this region were predicted to explode as Type Ib supernovae. This is because the rotating models lose more mass early in their evolution, leading to CO core masses below $40$~\msol~at all initial masses in this region.
\section{Stellar populations}
\label{section:pop}
In this section, the above results are placed in the context of a population of massive stars, weighted according to two different initial mass functions (IMFs). The distributions were calculated such that there is one star with $M_{\mathrm{ini}}=500~M_{\odot}$~in the Salpeter IMF distribution, with a total population of 90166 massive stars. This was calculated using Eq. (\ref{equation:imf}), with an exponent of $\alpha = 2.35$ for stars with $M > 0.5~M_{\odot}$~\citep{salpeter1955}. 
    \begin{equation}
        \frac{dN}{dM} = M^{-\alpha}
        \label{equation:imf}
    \end{equation}
 Then, a top-heavy mass distribution was calculated such that the population size remained the same as that calculated using the Salpter IMF. The 30 Doradus star-forming region in the Large Magellanic Cloud ($Z=0.006$) has been found to contain up to $32\%$ more stars with $M > 30$~\msol~than predicted by the Salpeter IMF. An exponent of $\alpha=1.90$ has been calculated based on spectroscopic observations of stars with mass ranging from $15$~\msol~to $200$~\msol. It is important to note that a significant proportion of the sample stars considered by \citet{schneider2018} were expected to be products of mass transfer in binary systems, and so the calculated IMF exponent may not be accurate for single stars. However, binary mass transfer also results in stars appearing younger than they are, and so these two effects may cancel each other out~\citep{schneider2013}. This highlights the uncertainty in such calculations, which is important to consider when drawing conclusions from IMF weighted proportions of remnant and supernova types. In this case, there are $\sim4$ stars with $M_{\mathrm{ini}}=500~M_{\odot}$, illustrating the top-heavy nature of this IMF when compared to the Salpeter IMF. 
\subsection{Fraction of massive stars per remnant type}
\label{section:fraction_remnant}
When considering the fraction of massive stars per remnant type, the NS and NS (BH) categories will be considered together; this combination is given as `Total NS' in the data tables presented in Appendix \ref{appendix:fraction_remnant}. Likewise, the BH and BH (NS) categories will be considered together; this combination is given as `Total BH'.
    \subsubsection{Impact of metallicity}
\begin{figure*}
	\includegraphics[width=0.8\textwidth]{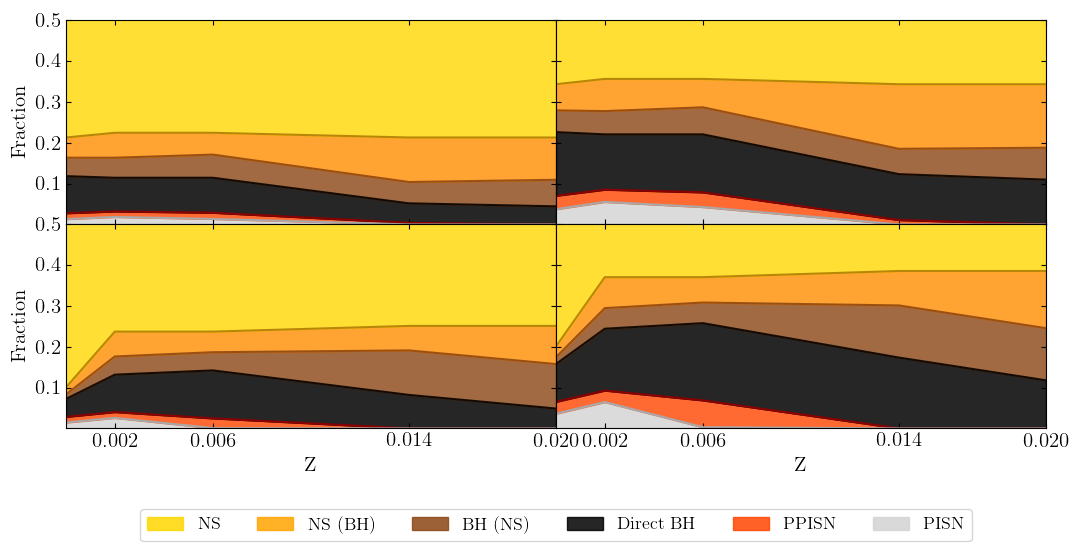}
    \caption{Fraction of non-rotating (top) and rotating (bottom) massive stars per remnant type, weighted by both the Salpeter IMF (left) and a top-heavy IMF (right).}
    \label{fig:fraction_rem}
\end{figure*}

    The distribution of remnants from non-rotating models weighted by the Salpeter IMF is considered when exploring the impact of metallicity, with results given in Table \ref{table:frac_rem_top} and also shown by the top left panel of Fig. \ref{fig:fraction_rem}. A significant proportion of massive stars are predicted to end their lives as NS at all metallicities considered. This is because massive stars with initial mass between $9-30$~\msol~are more heavily weighted than those with higher mass; most stars in this mass range are predicted to end their lives as NS, and so the fraction remains high across all metallicities. The fraction of massive stars predicted to end their lives as BH increases with metallicity until $Z=0.006$, then sharply decreases due to mass loss when considering solar and supersolar metallicities.
    The fraction of PPISN decreases as metallicity increases, such that very few are expected at solar metallicity, and none are expected at supersolar metallicity. This decrease is due to increased mass loss at higher metallicity, leading to smaller CO core masses that do not exceed the $M_{\mathrm{CO}}>40$~\msol~threshold for PPISN at all masses considered. The fraction of PISN is similar to that of PPISN from $Z=10^{-5}$ to $Z=0.006$, with a slight peak at $Z=0.002$. When $Z>0.006$, this fraction is 0 as no stars are predicted to have a CO core mass above the  $M_{\mathrm{CO}}>60$~\msol~threshold for PISN due to high levels of mass loss.

    Therefore, the main effect of metallicity on the fraction of massive stars per remnant type is due to the metallicity dependency of mass loss, given by Eq. (\ref{z_ml}). At low metallicity, there are more massive CO cores due to low levels of mass loss, increasing the fraction of (P)PISN and BH. At higher metallicity, increased levels of mass loss mean that the CO core mass does not exceed $40$~\msol~and so there are very few (P)PISN, if any at all. In addition, the high levels of mass loss mean that smaller CO core masses are more common, hence the increase in the fraction of NS at higher metallicities. 
    \subsubsection{Impact of rotation}
    The distribution of remnants from rotating models weighted by the Salpeter IMF is shown in the bottom left panel of Fig. \ref{fig:fraction_rem}. Results are given in Table \ref{table:frac_rem_top} in Appendix \ref{appendix} and the main differences are clear when comparing the top left and bottom left panels of Fig. \ref{fig:fraction_rem}. The majority of massive stars are still predicted to end their lives as NS, for the same reasons as outlined for non-rotating massive stars. One key difference between the rotating and non-rotating distribution of remnants is the significant increase in the fraction of NS at $Z=10^{-5}$; this is due to interactions between the hydrogen burning shell and helium burning core leading to a smaller CO core mass in rotating models. Massive stars within this mass range are heavily weighted by the IMF, hence this effect has a significant impact on the fraction of NS. 
    
    The effect of metallicity is different for the rotating case, due to the hydrogen burning shell effect and the increased impact of mass loss in rotating models. The fraction of BH also shows a strong dependence on metallicity. Rotation decreases the fraction of BH at $Z=10^{-5}$ significantly. This is because stars that would have massive enough CO cores to form BH if they were non-rotating experience the hydrogen shell effect, leading to smaller CO cores that lead to NS instead. 
    
    For $Z \geq 0.002$, the fraction of massive stars ending their lives as BH is higher for rotating stars than their non-rotating counterparts. This is due to rotation-induced mixing generally leading to larger core masses. The most significant difference is at solar metallicity, where the fraction of rotating stars expected to end their lives as BH is almost double that of non-rotating stars. This is because of the increase in failed supernova leading to BH at this metallicity, which has a significant impact on the fraction due to the IMF weighting placing more emphasis on stars with mass between $9-30$~\msol.
    
    Rotation increases the fraction of PPISN when metallicity is less than $Z=0.006$. Above this metallicity, no PPISN are predicted since the CO core mass of stars with solar and supersolar metallicity does not exceed 40~\msol~due to increased mass loss experienced by rotating stars. Similarly, the fraction of PISN is higher for rotating stars when $Z \leq 0.002$, but is almost zero at $Z=0.006$ as very few stars at this metallicity have a CO core mass exceeding 60~\msol.

    \subsubsection{Impact of using a top-heavy IMF}
    The fraction of massive stars per remnant type is qualitatively very similar when comparing results weighted by either the IMF
    (see right panels in Fig. \ref{fig:fraction_rem}).

Quantitatively, using a top-heavy IMF results in a smaller fraction of NS, and increased fractions of BH and (P)PISN. \resub{Additionally, using the top-heavy IMF means that a larger weighting is given to stars with mass between $30-100$~\msol, and in relative terms an even larger increase in weighting is given to VMSs. This means that the fraction of (P)PISN experiences the most noticeable increase (in relative terms) when considering the top-heavy IMF. Table \ref{table:frac_rem_top} shows that these fractions are almost three times higher than those calculated using the Salpeter IMF at SMC and LMC metallicities. It is important to note that despite a fraction of 0.066 of rotating massive stars expected to result in PPISN at LMC metallicity (calculated using the top-heavy IMF from \citet{schneider2018} based on an area of the LMC), there have been no confirmed observations of PPISN. Considering absolute fractions, the largest increase is seen for direct black holes, which occur within the initial mass range of $30-100$~\msol~(mass range more populated than VMSs) and can reach 0.06-0.08 at SMC and LMC metallicities.}

\subsection{Fraction of massive stars per supernova type}
   \subsubsection{Impact of metallicity}

\begin{figure*}
	\includegraphics[width=0.8\textwidth]{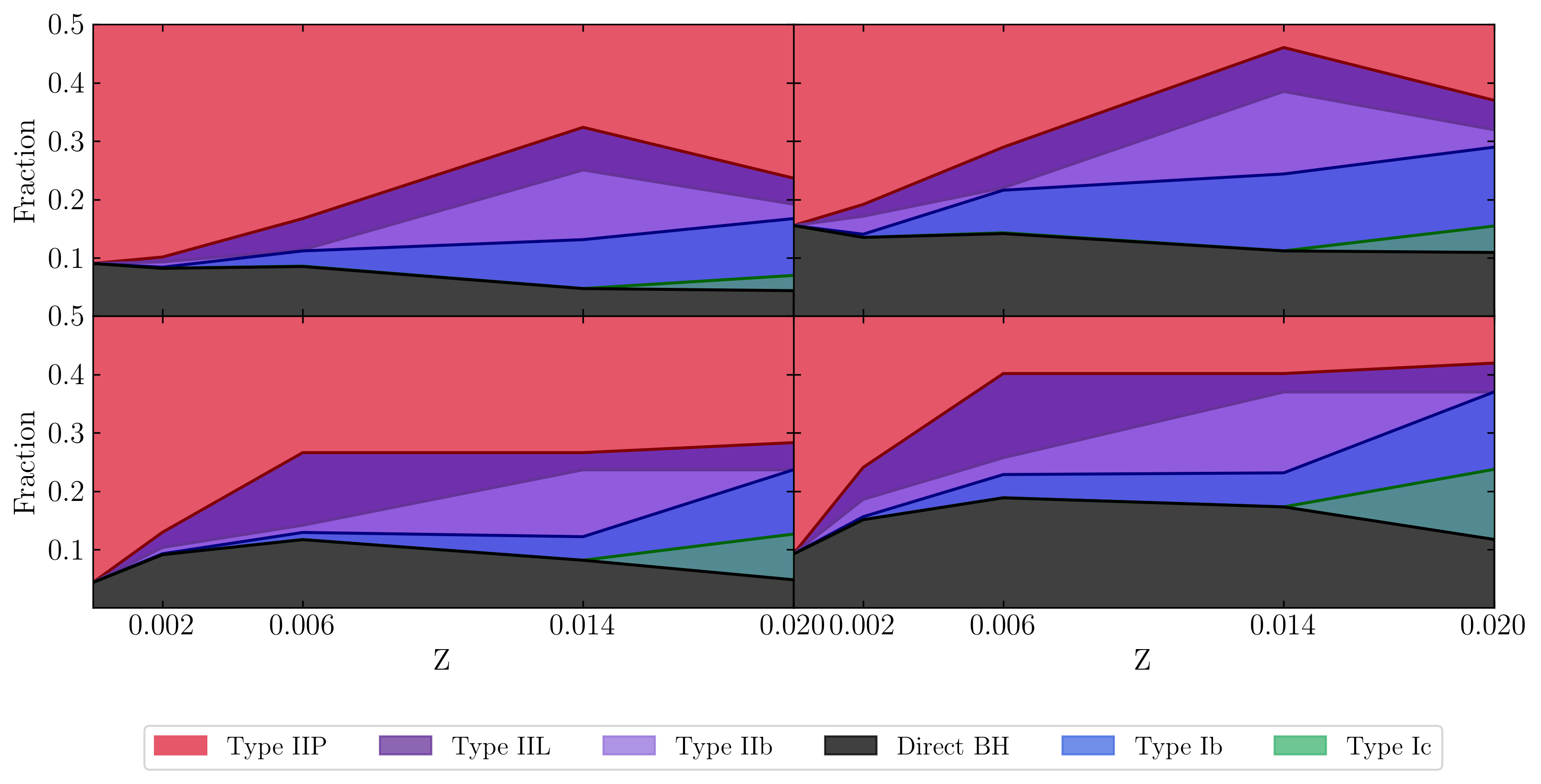}
    \caption{\resub{Fraction of non-rotating (top) and rotating (bottom) massive stars per supernova type, weighted by both the Salpeter IMF (left) and a top-heavy IMF (right)}}
    \label{fig:fraction_sn}
\end{figure*}

   The distribution of supernova type from non-rotating models weighted by the Salpeter IMF is considered when exploring the impact of metallicity, with results given in Table \ref{table:frac_sn_top} and also shown by the top left panel of Fig. \ref{fig:fraction_sn}. When $Z=10^{-5}$, almost all massive stars are expected to explode as Type IIP supernovae. No other types of supernovae are expected at this metallicity, with the remaining fraction expected to directly collapse to BH. The low rates of mass loss experienced by stars at this metallicity means that the envelope is rich in both hydrogen and helium across the range of initial masses considered in this work. As metallicity increases from $Z=10^{-5}$ to $Z=0.014$, the fraction of Type IIP supernovae decreases. Then, from $Z=0.014$ to $Z=0.02$, the fraction of Type IIP supernovae increases. This is because mass loss occurs later in the evolution of stars with initial mass below $25$~\msol~at solar metallicity when compared to supersolar metallicity. Note that models below $25$~\msol~at both metallicities have similar CO core masses and final masses, it is only the envelope mass which differs. This shows that they undergo a similar amount of mass loss throughout the whole evolution, but models at solar metallicity undergo more mass loss \textit{later} in the evolution. The fraction of massive stars predicted to explode as Type IIL supernovae is zero when $Z=10^{-5}$, increasing with metallicity until $Z=0.014$. This increase is due to increased levels of mass loss with metallicity, due to the dependence given by Eq. (\ref{z_ml}). The fraction of Type IIL supernovae then decreases from solar to supersolar metallicity, as more stars are expected to explode as Type IIP supernovae. 

   \resub{No stars at $Z=10^{-5}$ are expected to explode as Type IIb or Ib supernovae, since they all have envelopes rich in hydrogen. From $Z=0.002$ to $Z=0.014$, the fraction of massive stars expected to explode as Type IIb or Ib supernovae increases due to the increased rates of mass loss experienced by models at this metallicity. This fraction then decreases from solar to supersolar metallicity (due to stars losing mass earlier in their evolution). }The fraction of massive stars predicted to explode as Type Ic supernovae is very low, or zero, across all metallicities considered. At EMP, SMC and solar metallicities, no Type Ic supernovae are predicted. When $Z=0.006$, a small fraction is expected to explode as Type Ic supernovae, while at $Z=0.02$ this is slightly higher. These supernovae are predicted to be very rare for massive single stars, as very few have envelopes \resub{free of hydrogen and (almost) free of helium}. Those that do tend to have high initial mass, and so are not favourably weighted by the Salpeter IMF. Finally, the fraction of massive stars predicted to directly collapse to black holes is the same as discussed in Section \ref{section:fraction_remnant}. Hence, metallicity has a significant impact on supernova type, largely due to the metallicity dependence of mass loss.

        \subsubsection{Impact of rotation}
The distribution of \resub{supernova types} from rotating models weighted by the Salpeter IMF is shown in the bottom left panel of Fig. \ref{fig:fraction_sn}. Results are given in Table \ref{table:frac_sn_top} in Appendix \ref{appendix} and the main differences are clear when comparing the top left and bottom left panels of Fig. \ref{fig:fraction_sn}. When $Z=10^{-5}$, the fractions are very similar between the non-rotating and rotating case, with the only predicted outcomes being either a Type IIP supernova or direct collapse to a black hole. In the rotating case, the fraction of massive stars predicted to explode as Type IIP supernovae is higher than the same fraction in the non-rotating case. This is because of the interaction between the hydrogen burning shell and helium burning core in stars at this metallicity, with fewer rotating stars expected to directly collapse to BH when compared to non-rotating stars. The fraction of massive stars expected to explode as Type IIL supernovae is higher for rotating stars when $0.002 \leq Z \leq 0.006$, as more rotating stars partially lose their hydrogen envelope. This fraction is lower than that for non-rotating stars at solar metallicity due to the increased fraction of direct BH. At supersolar metallicity, this fraction is very similar between the non-rotating and rotating case, as shown by the top left and bottom left panels of Fig. \ref{fig:fraction_sn}. 
  
        \resub{Rotation does not really affect the fraction of type IIb nor the fraction of type Ib at low metallicities.
        The fraction of massive stars predicted to explode as Type Ib supernovae when $Z=0.006$ is slightly lower for rotating models, which is partly balanced by the corresponding increase in type IIb at that metallicity. 
        When $Z\geq 0.014$, the type Ib fraction is lower for rotating stars. }
        This is because more rotating stars are expected to directly collapse to BH, which would have exploded as Type Ib supernovae in the non-rotating case.
        The fraction of rotating massive stars expected to explode as Type Ic supernovae is zero, apart from at supersolar metallicity where it is more than double that in the non-rotating case. This is because more rotating stars are expected to be completely free of both hydrogen and helium due to the increased mass loss rates which they experience. Hence, the effect of rotation on supernova type is dominated by increased mass loss, as well as the increased number of stars expected to directly collapse to BH. 
        
        \subsubsection{Impact of using a top-heavy IMF}
        The fraction of massive stars per supernova type is qualitatively very similar when comparing results weighted by either the Salpeter or top-heavy IMF, with the distribution of supernova types weighted by a top-heavy IMF given by the rightmost panels of Fig. \ref{fig:fraction_sn}. Using the top-heavy IMF results in a decrease in the fraction of Type IIP supernovae, and an increase in the fraction of all other possibilities across the whole range of metallicities considered. This is because stars that explode as Type IIP supernovae generally have a lower initial mass (apart from when $Z=10^{-5}$), and these stars make up less of the stellar population calculated using the top-heavy IMF than the Salpeter IMF. 
        \resub{As discussed in Section \ref{section:fraction_remnant}, the largest increase in absolute fractions is seen for direct black holes. These increases are followed closely by increases in Type Ib fractions for non-rotating models at LMC and solar metallicities.
       The largest relative increase is found in type Ib fractions for the rotating SMC and LMC metallicity models, around three times higher than those calculated using the Salpeter IMF.}

        \subsection{Fraction of massive stars per supernova type expected to be PISN}
\begin{figure}
	\includegraphics[width=0.49\textwidth]{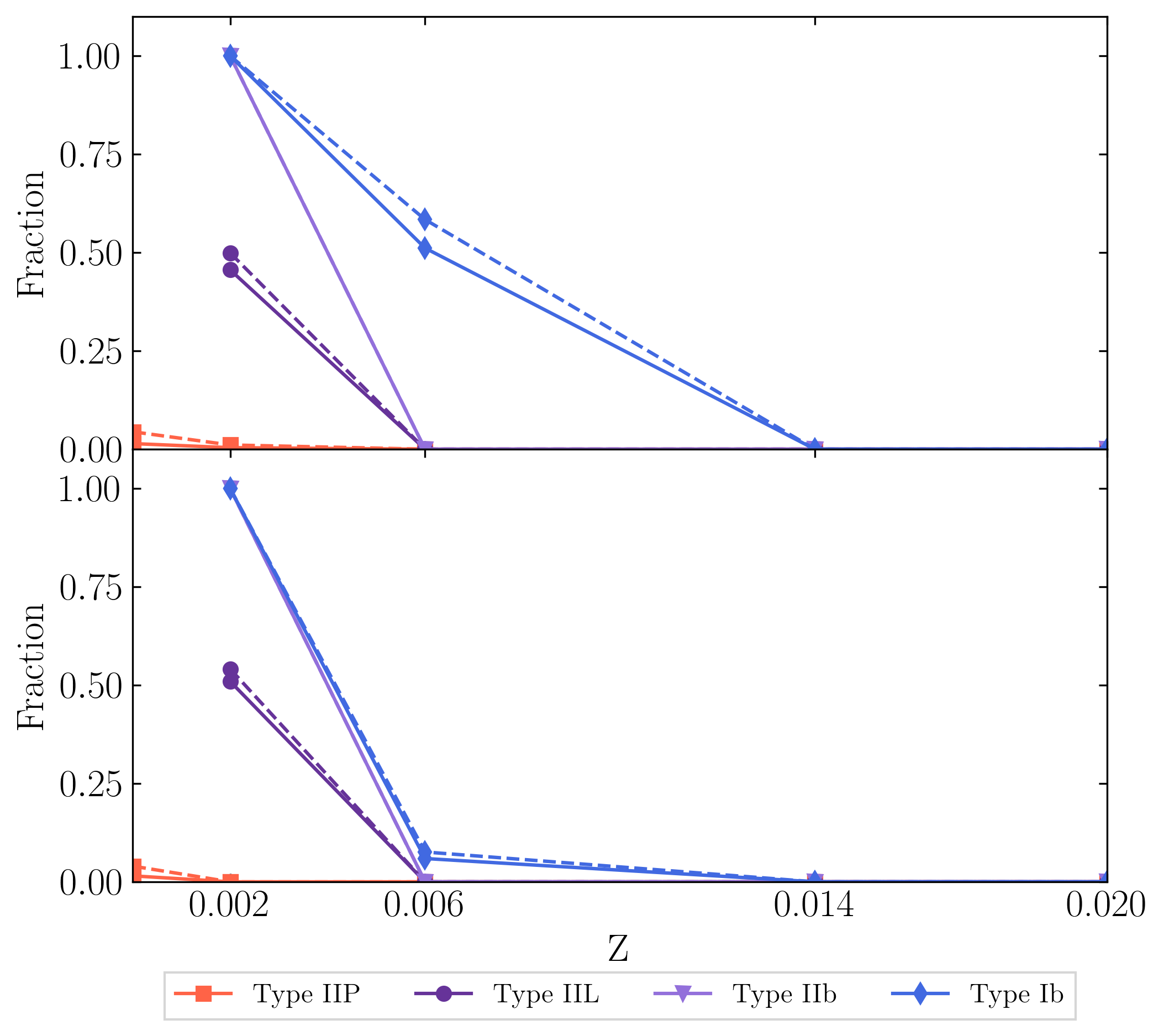}
    \caption{\resub{Fraction of non-rotating (top) and rotating (bottom) massive stars per supernova type that are expected to be PISN. Solid lines indicate those weighted using the Salpeter IMF and dashed lines for the top-heavy IMF.}}
    \label{fig:frac_pisn}
\end{figure}
Fig. \ref{fig:frac_pisn} shows, for each supernova type, the fraction, which is expected to be PISN (rather than core-collapse SN). It is important to note that no PISN are expected at solar or supersolar metallicities, for both the non-rotating and rotating case using both variations of the IMF. Firstly, the impact of metallicity will be considered, for the non-rotating case calculated using the Salpeter IMF. When $Z=10^{-5}$, the fraction of Type IIP supernova expected to be PISN is small, and decreases even further with increasing metallicity  reaching 0 when $Z=0.006$. This is because PISN have progenitors at low metallicity which are \resub{VMSs}, and progenitors of Type IIP supernovae are generally not VMSs when $Z>0.002$.
However, the fraction of Type IIL supernovae expected to be PISN is significantly higher and \textit{all} Type Ib supernovae at SMC metallicity are expected to be PISN considering single stars (see discussion about binarity in the next section). When $Z=0.006$, no Type IIP/L supernovae are expected to be PISN (the models predict more type Ib PPISN at $Z=0.006$), and the fraction of Type Ib supernovae expected to be PISN lower than when $Z=0.002$  (the models predict more type Ib PPISN at $Z=0.006$). Hence, the impact of metallicity depends on the supernova type considered.

The most significant effect of rotation on PISN types is the significant decrease in the fraction of Type Ib supernovae expected to be PISN in the rotating case. This is because PISN are only expected when $Z\leq0.006$ in the rotating case due to lower CO core masses than in the non-rotating case. Using the top-heavy IMF results in an increase in all of the fractions, other than those equal to either 0 or 1; the choice of IMF has a significant impact on the fraction of Type IIP supernovae expected to be PISN, and very little impact on that of Type IIL and Ib. 

\section{Discussion}
\label{discussion}
\subsection{Comparison to observations}

\resub{Table \ref{table:obs_sn} 
presents the proportions of different supernova types derived from large observational surveys published in  \citet{smith2011} and \citet{li2011}, which we can compare to the fractions predicted by the present models (Table\,\ref{table:frac_sn_top}). A detailed comparison is difficult as we do not know the distribution of metallicity of the observed SN and not even the average metallicity of the large samples (although the average is often considered to fall in between the LMC and solar metallicities). Nevertheless, we can still compare them in general terms. 
The observed fraction of Type IIP supernovae is lower than the fraction found in the models.
Only a top heavy IMF at not too low metallicities can reach the observed values.
The observational fraction of Type IIL supernovae can be reproduced by the models though model predictions vary a lot, especially for the models with rotation. 
The observed fraction of Type IIb supernovae is similar to the fraction found in the solar metallicity models but much higher than that found in low metallicity models. The observational fractions of Type Ib supernovae vary from \citet{smith2011} to \citet{li2011}, but can be explained by the fractions predicted in Section \ref{section:sn}, though the fractions calculated using the top-heavy IMF are higher than the observational range in Type Ib fraction.}

\resub{The most noticeable difference is for the fraction of Type Ic, 
where the observed fraction is much higher than the models predict (even when considering a top-heavy IMF).
This is in part compensated by the higher predicted fraction of Type Ib in the models and it is still not completely clear what differentiates a Type Ic from a Type Ib SN from the progenitor point of view as discussed in Sect.\,\ref{section:sn-types}. An alternative explanation is that Type Ic SN come mainly from stars that have experienced interactions with a binary companion. Binary interactions are indeed expected to be common for massive stars as discussed below. Even when considering binary stars, it remains a challenge to remove the helium rich layer and predict a Type Ic and it is not clear how much helium can be hidden in a Type Ic SN \citep[see e.\,g.][]{aguilera-dena2023,Jin2023}.
}

\begin{table}
\caption{
Observational fractions of core collapse supernova types from \citet{smith2011} and \citet{li2011}.}
\centering
\begin{tabular}{ccccccc}
\hline
  &  IIP &  IIL & \resub{IIb} & Ib &  Ic & \resub{Other} \\
\hline
\citet{smith2011} & 0.482 & 0.064 & \resub{0.106} & 0.071 & 0.149 & \resub{0.128} \\
\citet{li2011} & 0.521 & 0.074 & \resub{0.089} & 0.052 & 0.134 & \resub{0.129} \\
\end{tabular}
\label{table:obs_sn}
\end{table}

\subsubsection{Pair-instability supernovae}
\label{section:pisn}
PISN are usually expected to appear as superluminous supernovae with broad light curves \citep{gal2012}. 
Given that the IMF favours stars at the bottom of the PISN mass range, the majority of PISN, however, do not produce the extremely bright superluminous SN generally expected and are instead not much more luminous than core-collapse SN \citep[see e.\,g. Fig. 13 in ][]{gilmer2017}. Furthermore, they can evolve faster than generally expected from Type II superluminous SN as they often have lost their H-rich envelope as discussed above \citep[see e.\,g.][]{kozyreva2017}.
Even considering the brightest PISN from the top of the PISN mass range, PISN are difficult to find with optical transient surveys as they are expected to occur at high redshift, and so are faint in the optical wavelength range. For example, the deep optical survey with Subaru/Hyper Suprime-Cam (HSC), aimed at finding PISN candidates, has so far not found any \citep{moriya2021}. Instead, the use of near-infrared instruments is more suitable for searching for PISN as they are bright in the near-infrared range. In the next decade, several wide-field near-infrared instruments are planned, such as the near-infrared wide-field instrument on the recently launched Euclid satellite \citep{scaramella2022} and the Wide Field Instrument on the Nancy Grace Roman space telescope \citep{spergel2015}, which have the potential to find PISN candidates. However, there are currently no confirmed observations of PISN, and so no comparisons can be made in this case. It is important to note that there are, however, a few recent PISN and PPISN candidates \citep{angus2024, aamer2024, schulze2024, gomez2019}.

Despite the weak observational evidence for PISN, it is still important to consider the proportion of supernovae which may be PISN as they have very different nucleosynthetic yields, e.g. the odd-even effect is more significant for PISN than core-collapse supernovae, alongside higher abundances of Si, S, Ar and Ca due to the explosive oxygen burning \citep{kobayashi2015}. This has implications within the field of Galactic chemical evolution, where these nucleosynthetic yields are used within simulations of chemical enrichment processes. However, \citet{kobayashi2020} did not include effects from PISN in their Galactic chemical evolution models due to the lack of observational evidence. Hence, advancements in understanding whether PISN are likely to occur will be important if the observational data does become available. Additionally, the fraction of supernovae expected to be PISN is important when considering the distribution of black hole masses \citep{marchant2016, farmer2019, winch2024}, in particular the PI mass gap discussed in Section \ref{section:bh}.

\subsection{Effect of binarity}
Most massive stars form in multiples and interactions between massive binary stars are common as they tend to exist in close binary systems \citep{chini2012,sana2012,preibisch2001}. Due to this, it is important to consider the effect that binarity would have on the results presented in this study \citep[see][for a review on the various ways binarity affects the evolution of massive stars]{langer2012}. 

Mass transfer has a significant impact on the evolution of both stars involved, and so will influence their respective fates. If the mass transfer occurs before the end of core helium burning (Case A or B), it can change the mass of the donor star's convective core \resub{ \citep[see e.\,g.][]{laplace2021,schneider2021}.} This means that it would end core helium burning with a smaller CO core mass, and so this would affect the remnant type predicted in this work. It may possibly result in more NS, and reduce the fraction of massive stars expected to end their lives as PISN, as this requires that very high CO core masses are retained. Additionally, loss of the envelope means that the fraction of massive stars expected to explode as Type Ib/c supernovae would be higher when including effects due to mass transfer. On the other hand, the secondary star in the binary system would have a larger envelope mass, and so may retain more hydrogen/helium in the envelope than if it were a single star. Hence, the effects that mass transfer has on the fate of massive stars is complicated. It is important to note that \citet{de2002} found that including binary stars in models of galactic chemical evolution had only a moderate effect on yield, suggesting that considering only single stars may well still represent the average properties of massive stars.

In close binary systems, stars tend to synchronise their rotation to the orbital period, causing tidal mixing that dissipates the excess kinetic energy \citep{zahn1975, zahn1977,zahn2013}. These internal tides increase rotational mixing, and so increase the effects of rotation even further. Additionally, this leads to the star being very compact, as it is in quasi-chemical equilibrium throughout its evolution. This means that it is less likely to overfill its Roche lobe, and so may prevent any resulting mass transfer. Hence, the effects of mass transfer on the fate of massive stars may be less significant than described above. \resub{It is also worth noting that at low metallicities, stars expand less (see Fig.\,\ref{fig:hrd}) while at high metallicities the most massive stars lose so much mass that they also expand less \citep[][]{romagnolo2024}. Such limited expansion might affect or event suppress binary interactions \citep[][]{romagnolo2025}. }  
There are numerous possible intermediate products and end-points of the evolution of massive binary stars, highlighting the complexity of considering binary interactions when determining the fate of massive stars, and why it is beyond the scope of this work. See \citet{eldridge2017} for an example of how stellar evolution models can include effects due to binarity.

\section{Conclusions}
\label{conclusion}
In this work, we use the large grid of GENEC models, covering initial masses from 9 to 500~$M_{\odot}$~and metallicities ranging from $Z=10^{-5}$ to $0.02$, to study the impact of initial mass, metallicity and rotation on the fate of massive stars.
As initial mass is varied, different fates are found. At low initial mass, most massive stars are found to result in Type IIP supernovae and the formation of a neutron star. Then, as the initial mass increases, stars are less likely to successfully explode, and so are more likely to form black holes. \resub{VMS}s are expected to end their lives as either PPISN or PISN (at low metallicity), or direct collapse to black holes (at higher metallicity). Extremely metal poor stars with the highest initial masses are also expected to directly collapse to black holes. Hence, varying the initial mass has a significant impact on the fate of massive stars. 

This has also been explored through choice of IMF, considering different mass distributions and how this affects the results presented. The use of a top-heavy IMF allows for a closer match to observational data on supernova type than the use of the Salpeter IMF, possibly suggesting that the top-heavy IMF is a better match for the mass distribution of massive stars. However, the observational data does not exclude supernovae resulting from stars in binary systems, and so any comparisons should be treated with caution. 

As the metallicity is increased, the rates of mass loss on the main sequence also increase. This results in reduced CO core masses at higher metallicities, having various impacts on the remnant it leaves behind. In particular, \resub{VMS}s with low metallicity can end their lives as PISN and so leave behind no remnant. As long as the star does not become a RSG, increasing the metallicity also increases the rate of post-main sequence mass loss. This means that a significant proportion of the envelope may be lost, possibly resulting in Type Ib/c supernovae. Hence, metallicity has a significant impact on the final fate of massive stars as a whole, with both the remnant and supernova type showing a strong dependence on the metallicity. 

Rotation has also been shown to have a significant impact on the fate of massive stars. It has two main competing effects: increased mixing tends to increase the CO core mass, and increased mass loss tends to decrease the CO core mass and H/He envelope mass. In addition to these effects, an unusual boost in the hydrogen burning shell of extremely metal poor models leads to a decrease in the CO core mass, which was unexpected. The dominant effect of rotation differs depending on initial mass and metallicity. Hence, rotation has different effects on the fate of massive stars depending on the other parameters considered.

The findings of this work are summarised below:
\begin{itemize}
    \item The competing effects that rotation has on the evolution of massive stars results in it having a complex effect on both remnant and supernova type, depending on the initial mass and metallicity of the star.
    \item A pair-instability mass gap is expected from $\sim 90~M_{\odot}$ to 
    $\sim 150~M_{\odot}$ with 
    extremely metal poor stars with the highest initial masses expected to directly collapse to black holes with mass \textit{above} the pair-instability mass gap.
    \item Pair-instability supernovae and pulsation pair-instability supernovae are predicted at metallicities lower than solar, and initial masses greater than $100~M_{\odot}$. 
\end{itemize}

Given the above predictions concerning PISN, if upcoming near-infrared wide-field surveys do not find potential PISN candidates at high redshift, then the theoretical framework on which the predictions in this work are based may need to be re-evaluated although the lack of observed PISN might still be explained by the IMF favouring the less bright PISN as discussed above. More importantly, if there are detections of gravitational wave events involving merging black holes with masses within the predicted mass gap, this will seriously challenge stellar evolution theory or support the multiple merger origin of GW events in dense environments \citep[see e.\,g.][]{vaccaro2024,antonini2025}.

Understanding the fate of massive stars, and how it is impacted by various factors, has important applications across many different fields within astrophysics. The way in which massive stars die, both the type of supernova they produce and the type of compact remnant they leave behind, has significant effects on models of galactic chemical evolution, nucleosynthetic yields, gravitational wave astronomy, and also provides the necessary context for the interpretation of observational data. Since observations are generally limited to the surface properties of stars, the use of models and assumptions is important when considering the internal properties of massive stars. The fate of a massive star has its basis in both internal and surface properties, in particular the CO core mass and composition of the envelope. The grids of one-dimensional stellar evolution models used in this work have been calculated over the past decade, covering a wide range on initial masses, metallicities and considering models with and without rotation. This allows for analysis of how the properties, and the fate, of massive stars varies across this parameter space. 

There are still many uncertainties to consider when using stellar evolution models, such as the mass loss prescriptions used and the extent of their convective zones. Although the effects of binarity on the fate of massive stars have briefly been discussed here, future studies will benefit from including a quantitative analysis of these effects as well as the effects of magnetic fields on stellar evolution, though this will be challenging given the very large parameter space to be covered.
Considering the effect of a metallicity-dependent IMF and specific star formation rates are another avenues for further exploration to compare models to observations like SN type fractions. 
Using observational constraints and multi-dimensional stellar evolution models, these uncertainties can hopefully be reduced in the future as our understanding of stellar evolution continues to advance.

\section*{Acknowledgements}
\resub{This work builds upon the large grids of GENEC models computed by a large team and is based upon the MPhil thesis of K. Goodman \citep{chambers2024} analysing these grids}. RH also acknowledges the students who completed summer projects on this topic: Tom Griffiths, George Shrive and Jacob Shrive. RH and KN acknowledge support from the World Premier International Research Centre Initiative (WPI Initiative), MEXT, Japan. RH acknowledges the IReNA AccelNet Network of Networks (National Science Foundation, Grant No. OISE-1927130) and \resub{the Wolfson Foundation that part-funded the greenHPC facility at Keele}. GM acknowledges funding from the European Research Council (ERC) under the European Union’s Horizon 2020 research and innovation program (Grant No. 833925). NY acknowledges the Fundamental Research Grant Scheme grant number FRGS/1/2018/STG02/UM/01/2 and FRGS/1/2021/STG07/UM/02/4 under Ministry of Higher Education, Malaysia. This article is based upon work from the ChETEC COST Action (CA16117) and the European Union’s Horizon 2020 research and innovation programme (ChETEC-INFRA, Grant No. 101008324). \resub{KN acknowledge support from JSPS KAKENHI Grant Numbers JP20K04024, JP21H044pp, JP23K03452, and JP25K01046}.

\section*{Data Availability}

Table\,\ref{table:data} is available in csv format as supplementary material. Other data used or generated for this article can be shared upon reasonable request to the corresponding author.



\bibliographystyle{mnras}
\bibliography{references} 



\appendix
\onecolumn
\section{Data tables}
\subsection{Origin of the models}
\label{appendix:models}
\input{tables/models}
\subsection{Model data}
\label{appendix}
\input{tables/data}
\newpage
\subsection{Fraction of massive stars per remnant type}
\label{appendix:fraction_remnant}
\input{tables/fraction_remnant_new}
\subsection{Fraction of massive stars per supernova type}
\label{appendix:fraction_sn}
\input{tables/fraction_sn_new}
\bsp	
\label{lastpage}
\end{document}

%% file: tables/models.tex
\tablefirsthead{\toprule
    $Z$ & $v_{\mathrm{ini}}/v_{\mathrm{crit}}$  & \multicolumn{14}{c}{Initial mass of models}\\
            \midrule}

    \tablehead{\toprule
    $Z$ & $v_{\mathrm{ini}}/v_{\mathrm{crit}}$  & \multicolumn{14}{c}{Initial mass of models}\\
            \midrule}
    
\topcaption{List of models used and their origin where superscript 1 corresponds to \citet{sibony2024}, 2 to \citet{georgy2013}, 3 to \citet{yusof2013}, 4 to \citet{eggenberger2022}, 5 to \citet{martinet2023}, 6 to \citet{ekstrom2012}, 7 to \citet{yusof2022}. Models without superscript are unpublished and have been calculated for this work.} 
\begin{center}
        \begin{supertabular}[t]{cccccccccccccccc}
            $1\times10^{-5}$ & 0 & 9$^1$ & 12$^1$ & 15$^1$ & 20$^1$ & 25$^1$ & 30$^1$ & 40$^1$ & 60$^1$ & 85$^1$ & 120$^1$ & 150$^1$ & 200$^1$ & 300$^1$ & 500$^1$ \\
             &               0.4 & 9$^1$ & 12$^1$ & 15$^1$ & 20$^1$ & 25$^1$ & 30$^1$ & 40$^1$ & 60$^1$ & 85$^1$ & 120$^1$ & 150$^1$ & 200$^1$ & 300$^1$ & 500$^1$ \\
            0.002            & 0 & 9$^2$ & 12$^2$ & 15$^2$ & 20$^2$ & 25$^2$ & 32$^2$ & 40$^2$ & 60$^2$ & 85$^2$ & 120$^2$ & 150     & 200     & 300     &         \\
            &                0.4 & 9$^2$ & 12$^2$ & 15$^2$ & 20$^2$ & 25$^2$ & 32$^2$ & 40$^2$ & 60$^2$ & 85$^2$ & 120$^2$ & 150$^3$ & 200$^3$ & 300$^3$ &         \\
            0.006            & 0 & 9$^4$ & 12$^4$ & 15$^4$ & 20$^4$ & 25$^4$ & 32$^4$ & 40$^4$ & 60$^4$ & 85$^4$ & 120$^4$ & 150$^3$ & 180$^5$ & 300$^5$ & 500$^3$ \\
            &                0.4 & 9$^4$ & 12$^4$ & 15$^4$ & 20$^4$ & 25$^4$ & 32$^4$ & 40$^4$ & 60$^4$ & 85$^4$ & 120$^4$ & 150$^3$ & 200$^3$ & 300$^3$ & 500$^3$ \\
            0.014            & 0 & 9$^6$ & 12$^6$ & 15$^6$ & 20$^6$ & 25$^6$ & 32$^6$ & 40$^6$ & 60$^6$ & 85$^6$ & 120$^6$ & 150$^3$ & 200$^3$ & 300$^3$ & 500$^3$ \\
            &                0.4 & 9$^6$ & 12$^6$ & 15$^6$ & 20$^6$ & 25$^6$ & 32$^6$ & 40$^6$ & 60$^6$ & 85$^6$ & 120$^6$ & 150$^3$ & 200$^3$ & 300$^3$ & 500$^3$ \\
            0.02             & 0 & 9$^7$ & 12$^7$ & 15$^7$ & 20$^7$ & 25$^7$ & 32$^7$ & 40$^7$ & 60$^7$ & 85$^7$ & 120$^7$ & 150$^7$ & 200$^7$ & 300$^7$ & 500     \\
            &                0.4 & 9$^7$ & 12$^7$ & 15$^7$ & 20$^7$ & 25$^7$ & 32$^7$ & 40$^7$ & 60$^7$ & 85$^7$ & 120$^7$ & 150$^7$ & 200$^7$ & 300$^7$ &         \\
            \bottomrule
        \end{supertabular}
    \label{table:models}
\end{center}

%% file: tables/data.tex
\tablefirsthead{\toprule
    $M_{\mathrm{ini}}$ & Z & $v_{\mathrm{ini}}/v_{\mathrm{crit}}$ & $M_{\mathrm{fin}}$ & $M_{\mathrm{BH}}$ & $M_{\mathrm{\alpha}}$ & $M_{\mathrm{CO}}$ & $M_{\mathrm{H}}^{\mathrm{env}}$ & $M_{\mathrm{He}}^{\mathrm{env}}$ & Remnant type & SN type \\
            \midrule}

    \tablehead{\toprule
    $M_{\mathrm{ini}}$ & Z & $v_{\mathrm{ini}}/v_{\mathrm{crit}}$ & $M_{\mathrm{fin}}$ & $M_{\mathrm{BH}}$ & $M_{\mathrm{\alpha}}$ & $M_{\mathrm{CO}}$ & $M_{\mathrm{H}}^{\mathrm{env}}$ & $M_{\mathrm{He}}^{\mathrm{env}}$ & Remnant type & SN type \\
            \midrule}
            
    \topcaption{Initial mass ($M_{\mathrm{ini}}$), final total mass ($M_{\mathrm{fin}}$), defined as the total mass at the end of core helium burning, remnant mass in the case of black hole formation, 
$M_{\mathrm{BH}}$, helium core mass ($M_{\mathrm{\alpha}}$), CO core mass ($M_{\mathrm{CO}}$), H envelope mass ($M_{\mathrm{H}}^{\mathrm{env}}$), He envelope mass ($M_{\mathrm{He}}^{\mathrm{env}}$), followed by remnant and SN type of the models. The full table is available as supplementary material in csv format.} 
   \begin{center}
        \begin{supertabular}[t]{ccccccccccc}
            9   & $10^{-5}$ & 0   & 9.00   & & 2.46   & 1.02   & 4.74  & 3.11   & NS & Type IIP \\
            9   & $10^{-5}$ & 0.4 & 9.00   & & 2.68   & 1.12   & 4.20  & 3.51   & NS & Type IIP \\
            12  & $10^{-5}$ & 0   & 12.00  & & 3.36   & 1.58   & 6.04  & 4.23   & NS & Type IIP \\
            12  & $10^{-5}$ & 0.4 & 12.00  & & 3.46   & 1.59   & 5.33  & 4.81   & NS & Type IIP \\
            15  & $10^{-5}$ & 0   & 14.99  & & 4.36   & 2.28   & 7.18  & 5.36   & NS & Type IIP \\
            15  & $10^{-5}$ & 0.4 & 14.99  & & 5.20   & 2.63   & 5.80  & 5.48   & NS & Type IIP \\
            20  & $10^{-5}$ & 0   & 20.00  & & 6.18   & 3.75   & 8.82  & 7.24   & NS & Type IIP \\
            20  & $10^{-5}$ & 0.4 & 20.00  & & 5.47   & 2.98   & 6.82  & 8.27   & NS & Type IIP \\
            25  & $10^{-5}$ & 0   & 25.00  & & 8.32   & 5.55   & 10.18 & 9.07   & NS & Type IIP \\
            25  & $10^{-5}$ & 0.4 & 25.00  & & 8.50   & 5.52   & 7.92  & 10.48  & NS & Type IIP \\
            30  & $10^{-5}$ & 0   & 29.99  & (7.57) & 10.67  & 7.57   & 11.34 & 10.84  & BH (NS) & Type IIP \\
            30  & $10^{-5}$ & 0.4 & 22.84  & & 7.67   & 4.81   & 4.23  & 10.68  & NS & Type IIP \\
            40  & $10^{-5}$ & 0   & 39.99  & (11.67) & 15.38  & 11.67  & 13.37 & 14.60  & NS (BH) & Type IIP \\
            40  & $10^{-5}$ & 0.4 & 39.84  & & 5.47   & 3.14   & 8.49  & 20.58  & NS & Type IIP \\
            60  & $10^{-5}$ & 0   & 60.00  & 60.00 & 25.73  & 20.94  & 17.18 & 21.27  & BH & Direct BH \\
            60  & $10^{-5}$ & 0.4 & 56.61  & 56.61 & 17.46  & 13.42  & 10.58 & 24.36  & BH & Direct BH \\
            85  & $10^{-5}$ & 0   & 84.99  & 84.99 & 37.42  & 31.63  & 21.66 & 30.69  & BH & Direct BH \\
            85  & $10^{-5}$ & 0.4 & 57.22  & 57.22 & 44.28  & 37.36  & 2.13  & 14.67  & BH & Direct BH \\
            120 & $10^{-5}$ & 0   & 98.45  & 47.27 & 54.20  & 48.00  & 12.86 & 35.28  & PPISN & Type IIP \\
            120 & $10^{-5}$ & 0.4 & 86.62  & 27.07 & 68.53  & 59.50  & 2.88  & 20.02  & PPISN & Type IIP \\
            150 & $10^{-5}$ & 0   & 128.59 & 28.81 & 67.52  & 58.88  & 17.59 & 49.96  & PPISN & Type IIP \\
            150 & $10^{-5}$ & 0.4 & 131.84 & & 94.72  & 86.62  & 7.43  & 37.03  & PISN & Type IIP \\
            200 & $10^{-5}$ & 0   & 168.21 & & 93.16  & 83.24  & 19.43 & 64.24  & PISN & Type IIP \\
            200 & $10^{-5}$ & 0.4 & 158.24 & & 134.54 & 123.50 & 2.90  & 27.29  & PISN & Type IIP \\
            300 & $10^{-5}$ & 0   & 245.66 & & 142.38 & 117.67 & 22.37 & 89.36  & PISN & Type IIP \\
            300 & $10^{-5}$ & 0.4 & 282.38 & 282.38 & 266.28 & 247.63 & 1.06  & 22.14  & BH & Direct BH \\
            500 & $10^{-5}$ & 0   & 465.80 & 465.80 & 252.54 & 235.23 & 24.49 & 198.12 & BH & Direct BH \\
            500 & $10^{-5}$ & 0.4 & 462.66 & 462.66 & 433.15 & 404.05 & 1.62  & 36.09  & BH & Direct BH \\
            \midrule
            9   & $0.002$ & 0   & 8.89   & & 2.49   & 1.00   & 4.65 & 3.09  & NS & Type IIP \\
            9   & $0.002$ & 0.4 & 8.98   & & 2.85   & 1.14   & 4.14 & 3.39  & NS & Type IIP \\
            12  & $0.002$ & 0   & 11.91  & & 3.60   & 1.73   & 5.88 & 4.03  & NS & Type IIP \\
            12  & $0.002$ & 0.4 & 11.87  & & 3.84   & 1.80   & 5.09 & 4.79  & NS & Type IIP \\
            15  & $0.002$ & 0   & 14.78  & & 4.60   & 2.41   & 6.91 & 5.27  & NS & Type IIP \\
            15  & $0.002$ & 0.4 & 14.77  & & 4.97   & 2.59   & 5.86 & 5.95  & NS & Type IIP \\
            20  & $0.002$ & 0   & 19.70  & & 6.53   & 4.00   & 8.44 & 7.01  & NS & Type IIP \\
            20  & $0.002$ & 0.4 & 19.25  & & 7.09   & 4.24   & 6.65 & 7.63  & NS & Type IIP \\
            25  & $0.002$ & 0   & 24.40  & & 8.47   & 5.66   & 9.52 & 8.92  & NS & Type IIP \\
            25  & $0.002$ & 0.4 & 22.75  & (6.25) & 9.50   & 6.25   & 6.38 & 9.11  & BH (NS) & Type IIP \\
            32  & $0.002$ & 0   & 23.90  & (8.59) & 11.77  & 8.59   & 5.64 & 9.31  & NS (BH) & Type IIP \\
            32  & $0.002$ & 0.4 & 24.67  & (9.36) & 13.03  & 9.36   & 4.10 & 9.97  & NS (BH) & Type IIP \\
            40  & $0.002$ & 0   & 29.55  & (11.39) & 14.92  & 11.39  & 5.74 & 11.96 & NS (BH) & Type IIP \\
            40  & $0.002$ & 0.4 & 29.14  & 29.14 & 17.48  & 13.37  & 4.05 & 10.32 & BH & Direct BH \\
            60  & $0.002$ & 0   & 37.91  & 37.91 & 25.19  & 20.43  & 4.08 & 12.58 & BH & Direct BH \\
            60  & $0.002$ & 0.4 & 39.44  & 39.44 & 31.38  & 26.70  & 4.05 & 10.31 & BH & Direct BH \\
            85  & $0.002$ & 0   & 48.75  & 48.75 & 37.53  & 32.26  & 3.00 & 12.12 & BH & Direct BH \\
            85  & $0.002$ & 0.4 & 51.34  & 43.51 & 49.47  & 43.59  & 1.68 & 10.21 & PPISN & Type IIL \\
            120 & $0.002$ & 0   & 64.15  & 41.24 & 55.65  & 49.57  & 1.75 & 11.56 & PPISN & Type IIL \\
            120 & $0.002$ & 0.4 & 86.09  & & 84.78  & 76.61  & 1.67 & 10.20 & PISN & Type IIL \\
            150 & $0.002$ & 0   & 84.97  & & 70.91  & 63.39  & 2.69 & 16.78 & PISN & Type IIP \\
            150 & $0.002$ & 0.4 & 106.68 & & 106.68 & 93.78  & 0.27 & 5.02  & PISN & \resub{Type IIb} \\
            200 & $0.002$ & 0   & 105.92 & & 98.86  & 68.49  & 0.02 & 8.18  & PISN & Type Ib \\
            200 & $0.002$ & 0.4 & 129.34 & & 129.34 & 116.66 & 0.06 & 4.30  & PISN & \resub{Type IIb} \\
            300 & $0.002$ & 0   & 164.44 & & 158.59 & 116.57 & 0.08 & 31.35 & PISN & \resub{Type IIb} \\
            300 & $0.002$ & 0.4 & 149.78 & 149.78 & 149.78 & 134.10 & 0.00 & 3.18  & BH & Direct BH \\
            \midrule
            9   & $0.006$ & 0   & 8.65  & & 2.27  & 0.88  & 4.57 & 2.93 & NS & Type IIP \\
            9   & $0.006$ & 0.4 & 8.80  & & 2.56  & 1.00  & 4.16 & 3.44 & NS & Type IIP \\
            12  & $0.006$ & 0   & 10.95 & & 3.26  & 1.47  & 5.30 & 3.86 & NS & Type IIP \\
            12  & $0.006$ & 0.4 & 11.25 & & 3.81  & 1.77  & 4.68 & 4.11 & NS & Type IIP \\
            15  & $0.006$ & 0   & 14.00 & & 4.46  & 2.30  & 6.35 & 5.03 & NS & Type IIP \\
            15  & $0.006$ & 0.4 & 14.15 & & 5.05  & 2.68  & 5.43 & 5.58 & NS & Type IIP \\
            20  & $0.006$ & 0   & 15.33 & & 6.46  & 3.94  & 5.19 & 5.68 & NS & Type IIP \\
            20  & $0.006$ & 0.4 & 13.80 & & 7.13  & 4.33  & 3.04 & 5.58 & NS & Type IIP \\
            25  & $0.006$ & 0   & 12.04 & & 8.47  & 5.70  & 5.17 & 5.67 & NS & Type IIP \\
            25  & $0.006$ & 0.4 & 12.42 & (6.28) & 9.35  & 6.28  & 1.18 & 4.09 & BH (NS) & Type IIL \\
            32  & $0.006$ & 0   & 12.51 & (8.37) & 11.43 & 8.37  & 1.48 & 4.51 & NS (BH) & Type IIL \\
            32  & $0.006$ & 0.4 & 13.99 & (9.91) & 13.43 & 9.91  & 1.18 & 4.08 & NS (BH) & Type IIL \\
            40  & $0.006$ & 0   & 15.55 & (11.53) & 14.95 & 11.53 & 1.47 & 4.50 & NS (BH) & Type IIL \\
            40  & $0.006$ & 0.4 & 19.41 & 19.41 & 18.89 & 14.89 & 0.09 & 2.36 & BH & Direct BH \\
            60  & $0.006$ & 0   & 22.92 & 22.92 & 22.92 & 18.23 & 0.39 & 3.13 & BH & Direct BH \\
            60  & $0.006$ & 0.4 & 33.03 & 33.03 & 33.03 & 28.25 & 0.09 & 2.35 & BH & Direct BH \\
            85  & $0.006$ & 0   & 31.53 & 31.53 & 31.53 & 26.15 & 0.16 & 2.73 & BH & Direct BH \\
            85  & $0.006$ & 0.4 & 35.93 & 35.93 & 35.93 & 29.93 & 0.09 & 2.35 & BH & Direct BH \\
            120 & $0.006$ & 0   & 54.62 & 41.94 & 54.62 & 47.27 & 0.00 & 0.48 & PPISN & Type Ic \\
            120 & $0.006$ & 0.4 & 52.58 & 42.60 & 52.58 & 45.08 & 0.08 & 2.32 & PPISN & \resub{Type IIb} \\
            150 & $0.006$ & 0   & 59.68 & 37.76 & 59.68 & 51.72 & 0.00 & 0.79 & PPISN & Type Ib \\
            150 & $0.006$ & 0.4 & 45.67 & 45.67 & 45.67 & 38.70 & 0.08 & 2.31 & BH & Direct BH \\
            180 & $0.006$ & 0   & 71.06 & & 71.06 & 63.49 & 0.00 & 1.19 & PISN & Type Ib \\
            200 & $0.006$ & 0.4 & 51.11 & 42.52 & 51.11 & 43.55 & 0.00 & 1.41 & PPISN & Type Ib \\
            300 & $0.006$ & 0   & 91.35 & & 91.35 & 82.30 & 0.00 & 1.13 & PISN & Type Ib \\
            300 & $0.006$ & 0.4 & 54.14 & 42.30 & 54.14 & 46.45 & 0.00 & 0.88 & PPISN & Type Ib \\
            500 & $0.006$ & 0   & 94.68 & & 94.68 & 82.18 & 0.00 & 1.68 & PISN & Type Ib \\
            500 & $0.006$ & 0.4 & 74.89 & & 74.89 & 65.37 & 0.00 & 1.27 & PISN & Type Ib \\
            \midrule
            9   & $0.014$ & 0   & 8.80  & & 2.21  & 0.83  & 4.62 & 3.07 & NS & Type IIP \\
            9   & $0.014$ & 0.4 & 8.58  & & 2.92  & 1.18  & 3.61 & 3.34 & NS & Type IIP \\
            12  & $0.014$ & 0   & 11.36 & & 2.95  & 1.26  & 5.68 & 3.91 & NS & Type IIP \\
            12  & $0.014$ & 0.4 & 10.31 & & 3.83  & 1.83  & 4.08 & 3.56 & NS & Type IIP \\
            15  & $0.014$ & 0   & 13.34 & & 4.20  & 2.11  & 5.87 & 5.04 & NS & Type IIP \\
            15  & $0.014$ & 0.4 & 11.19 & & 5.01  & 2.69  & 3.38 & 4.01 & NS & Type IIP \\
            20  & $0.014$ & 0   & 9.02  & & 6.14  & 3.67  & 1.33 & 3.67 & NS & Type IIL \\
            20  & $0.014$ & 0.4 & 7.55  & & 7.04  & 4.36  & 3.38 & 4.01 & NS & Type IIP \\
            25  & $0.014$ & 0   & 8.77  & & 8.06  & 5.34  & 0.28 & 2.57 & NS & \resub{Type IIb} \\
            25  & $0.014$ & 0.4 & 9.91  & (6.54) & 9.59  & 6.54  & 0.15 & 2.33 & BH (NS) & \resub{Type IIb} \\
            32  & $0.014$ & 0   & 11.17 & (7.72) & 10.74 & 7.72  & 0.12 & 2.55 & BH (NS) & \resub{Type IIb} \\
            32  & $0.014$ & 0.4 & 10.21 & (7.05) & 10.21 & 7.05  & 0.15 & 2.33 & BH (NS) & \resub{Type IIb} \\
            40  & $0.014$ & 0   & 13.92 & (10.40) & 13.86 & 10.40 & 0.00 & 2.23 & NS (BH) & Type Ib \\
            40  & $0.014$ & 0.4 & 12.41 & (9.01) & 12.41 & 9.01  & 0.03 & 2.01 & NS (BH) & Type Ib \\
            60  & $0.014$ & 0   & 12.57 & (9.18) & 12.57 & 9.18  & 0.00 & 0.52 & NS (BH) & Type Ib \\
            60  & $0.014$ & 0.4 & 18.07 & 18.07 & 18.07 & 13.97 & 0.00 & 0.51 & BH & Direct BH \\
            85  & $0.014$ & 0   & 18.72 & 18.72 & 18.72 & 14.67 & 0.00 & 0.57 & BH & Direct BH \\
            85  & $0.014$ & 0.4 & 26.49 & 26.49 & 26.49 & 21.47 & 0.00 & 0.54 & BH & Direct BH \\
            120 & $0.014$ & 0   & 31.00 & 31.00 & 31.00 & 25.64 & 0.00 & 0.71 & BH & Direct BH \\
            120 & $0.014$ & 0.4 & 19.12 & 19.12 & 19.12 & 14.87 & 0.00 & 0.60 & BH & Direct BH \\
            150 & $0.014$ & 0   & 41.26 & 41.26 & 41.26 & 34.96 & 0.00 & 0.87 & BH & Direct BH \\
            150 & $0.014$ & 0.4 & 20.31 & 20.31 & 20.31 & 16.10 & 0.00 & 0.68 & BH & Direct BH \\
            200 & $0.014$ & 0   & 49.42 & 41.44 & 49.42 & 42.50 & 0.00 & 0.82 & PPISN & Type Ib \\
            200 & $0.014$ & 0.4 & 22.01 & 22.01 & 22.01 & 17.57 & 0.00 & 0.60 & BH & Direct BH \\
            300 & $0.014$ & 0   & 38.24 & 38.24 & 38.24 & 32.17 & 0.00 & 0.84 & BH & Direct BH \\
            300 & $0.014$ & 0.4 & 24.01 & 24.01 & 24.01 & 19.43 & 0.00 & 0.63 & BH & Direct BH \\
            500 & $0.014$ & 0   & 29.84 & 29.84 & 29.84 & 24.21 & 0.00 & 0.82 & BH & Direct BH \\
            500 & $0.014$ & 0.4 & 25.91 & 25.91 & 25.91 & 21.05 & 0.00 & 0.70 & BH & Direct BH \\
            \midrule
             9  & $0.02$ & 0   & 8.80  & & 1.21  & 1.14  & 4.55 & 2.95 & NS & Type IIP \\
             9  & $0.02$ & 0.4 & 8.74  & & 1.83  & 1.31  & 4.05 & 3.12 & NS & Type IIP \\
            12  & $0.02$ & 0   & 11.56 & & 2.98  & 1.58  & 4.55 & 2.95 & NS & Type IIP\\
            12  & $0.02$ & 0.4 & 10.36 & & 3.68  & 2.14  & 4.05 & 3.12 & NS & Type IIP \\
            15  & $0.02$ & 0   & 13.09 & & 4.09  & 2.24  & 5.71 & 4.01 & NS &Type IIP \\
            15  & $0.02$ & 0.4 & 10.83 & & 5.22  & 3.09  & 4.08 & 3.50 & NS & Type IIP \\
            20  & $0.02$ & 0   & 8.45  & & 6.03  & 3.68  & 5.63 & 4.83 & NS & Type IIP \\
            20  & $0.02$ & 0.4 & 7.27  & & 7.14  & 4.66  & 3.15 & 3.70 & NS & Type IIP \\
            25  & $0.02$ & 0   & 8.04  & & 8.04  & 5.37  & 1.12 & 3.25 & NS & Type IIL \\
            25  & $0.02$ & 0.4 & 9.08  & (6.67) & 9.08  & 6.67  & 0.01 & 1.22 & BH (NS) & Type Ib \\
            32  & $0.02$ & 0   & 10.71 & (7.77) & 10.71 & 7.77  & 0.00 & 1.92 & BH (NS) & Type Ib \\
            32  & $0.02$ & 0.4 & 9.80  & (7.16) & 9.80  & 7.16  & 0.00 & 1.07 & BH (NS) & Type Ib \\
            40  & $0.02$ & 0   & 11.33 & (8.64) & 11.33 & 8.64  & 0.00 & 2.22 & NS (BH) & Type Ib \\
            40  & $0.02$ & 0.4 & 11.63 & (8.97) & 11.63 & 8.97  & 0.00 & 0.32 & NS (BH) & Type Ic \\
            60  & $0.02$ & 0   & 10.77 & (8.24) & 10.77 & 8.24  & 0.00 & 0.33 & NS (BH) & Type Ic \\
            60  & $0.02$ & 0.4 & 12.87 & (9.93) & 12.87 & 9.93  & 0.00 & 0.40 & NS (BH) & Type Ic \\
            85  & $0.02$ & 0   & 16.21 & 16.21 & 16.21 & 12.91 & 0.00 & 0.38 & BH & Direct BH \\
            85  & $0.02$ & 0.4 & 16.64 & 16.64 & 16.64 & 13.25 & 0.00 & 0.39 & BH & Direct BH \\
            120 & $0.02$ & 0   & 23.40 & 23.40 & 23.40 & 19.15 & 0.00 & 0.43 & BH & Direct BH \\
            120 & $0.02$ & 0.4 & 22.26 & 22.26 & 22.26 & 18.05 & 0.00 & 0.42 & BH & Direct BH \\
            150 & $0.02$ & 0   & 30.92 & 30.92 & 30.92 & 26.07 & 0.00 & 0.49 & BH & Direct BH \\
            200 & $0.02$ & 0   & 35.65 & 35.65 & 35.65 & 30.02 & 0.00 & 0.63 & BH & Direct BH \\
            200 & $0.02$ & 0.4 & 34.64 & 34.64 & 34.64 & 29.09 & 0.00 & 0.48 & BH & Direct BH \\
            300 & $0.02$ & 0   & 22.23 & 22.23 & 22.23 & 18.08 & 0.00 & 0.69 & BH & Direct BH \\
            300 & $0.02$ & 0.4 & 25.24 & 25.24 & 25.24 & 20.62 & 0.00 & 0.67 & BH & Direct BH \\
            500 & $0.02$ & 0   & 25.55 & 25.55 & 25.55 & 20.31 & 0.00 & 0.67 & BH & Direct BH \\
            500 & $0.02$ & 0.4 & 25.24 & 25.24 & 25.24 & 20.62 & 0.00 & 0.67 & BH & Direct BH \\
            \bottomrule
        \end{supertabular}
        \label{table:data}
\end{center}

%% file: tables/fraction_remnant_new.tex
 \tablehead{\toprule Z & $v_{\mathrm{ini}}/v_{\mathrm{crit}}$ & $\alpha$ & NS & NS (BH) & Total NS & BH (NS) & BH & Total BH & PPISN & PISN \\ \midrule} 

\topcaption{Fraction of massive stars per remnant type, calculated using IMFs from \citet{salpeter1955} with $\alpha=2.35$ and \citet{schneider2018} with $\alpha=1.9$.}

\begin{center}
\begin{supertabular}{ccccccccccc}
$10^{-5}$ & 0 & 2.35 & 0.787 & 0.049 & 0.836 & 0.045  & 0.091 & 0.136 & 0.014 & 0.013 \\
0.002     & 0 & 2.35 & 0.776 & 0.061 & 0.837 & 0.049  & 0.082 & 0.131 & 0.013 & 0.018 \\
0.006     & 0 & 2.35 & 0.776 & 0.053 & 0.829 & 0.057  & 0.086 & 0.143 & 0.015 & 0.013 \\
0.014     & 0 & 2.35 & 0.787 & 0.109 & 0.896 & 0.052  & 0.048 & 0.100 & 0.004 & 0 \\
0.02      & 0 & 2.35 & 0.787 & 0.103 & 0.890 & 0.065  & 0.044 & 0.109 & 0     & 0 \\
\midrule
 $10^{-5}$ & 0.4 & 2.35 & 0.900 & 0.017 & 0.917 & 0.011 &  0.044 & 0.054 & 0.014 & 0.014 \\
 0.002     & 0.4 & 2.35 & 0.763 & 0.061 & 0.824 & 0.044 &  0.092 & 0.135 & 0.014 & 0.026 \\
 0.006     & 0.4 & 2.35 & 0.763 & 0.050 & 0.813 & 0.044 &  0.117 & 0.161 & 0.024 & 0.001 \\
 0.014     & 0.4 & 2.35 & 0.749 & 0.060 & 0.809 & 0.109 &  0.082 & 0.191 & 0 & 0 \\
 0.02      & 0.4 & 2.35 & 0.749 & 0.093 & 0.842 & 0.109 &  0.049 & 0.158 & 0 & 0 \\
\midrule
$10^{-5}$ & 0 & 1.9 & 0.657 & 0.064 & 0.721 & 0.053 & 0.156 & 0.209 & 0.033 & 0.037 \\
0.002     & 0 & 1.9 & 0.644 & 0.079 & 0.723 & 0.057 & 0.135 & 0.192 & 0.030 & 0.055 \\
0.006     & 0 & 1.9 & 0.644 & 0.069 & 0.713 & 0.066 & 0.142 & 0.208 & 0.036 & 0.043 \\
0.014     & 0 & 1.9 & 0.657 & 0.158 & 0.815 & 0.062 & 0.112 & 0.174 & 0.011 & 0 \\
0.02      & 0 & 1.9 & 0.657 & 0.155 & 0.812 & 0.078 & 0.110 & 0.188 & 0     & 0 \\
\midrule
 $10^{-5}$ & 0.4 & 1.9 & 0.799 & 0.026 & 0.825 & 0.016 & 0.093 & 0.109 & 0.029 & 0.036 \\
 0.002     & 0.4 & 1.9 & 0.630 & 0.076 & 0.706 & 0.050 & 0.151 & 0.201 & 0.028 & 0.065 \\
 0.006     & 0.4 & 1.9 & 0.630 & 0.062 & 0.692 & 0.050 & 0.189 & 0.239 & 0.066 &  0.003 \\
 0.014     & 0.4 & 1.9 & 0.615 & 0.084 & 0.699 & 0.128 & 0.173 & 0.301 & 0     & 0 \\
 0.02      & 0.4 & 1.9 & 0.615 & 0.140 & 0.755 & 0.128 & 0.118 & 0.246 & 0     & 0  \\
 \bottomrule
\end{supertabular}
\label{table:frac_rem_top}
\end{center}

%% file: tables/fraction_sn_new.tex
 \tablehead{\toprule   Z & $v_{\mathrm{ini}}/v_{\mathrm{crit}}$ & $\alpha$ & Type IIP & Type IIL & \resub{Type IIb} & Type Ib & Type Ic & Direct BH \\ \midrule} 

\topcaption{\resub{Fraction of massive stars per supernova type, calculated using IMFs from \citet{salpeter1955} with $\alpha=2.35$ and \citet{schneider2018} with $\alpha=1.9$.}}

\begin{center}
\begin{supertabular}{ccccccccc}
$10^{-5}$ & 0   & 2.35 & 0.909 & 0     & \resub{0    }  & 0     & 0     & 0.091 \\
0.002     & 0   & 2.35 & 0.898 & 0.008 & \resub{0.009} & 0.002 & 0     & 0.082 \\
0.006     & 0   & 2.35 & 0.833 & 0.053 & \resub{0.002} & 0.026 & 0.001 & 0.086 \\
0.014     & 0   & 2.35 & 0.676 & 0.073 & \resub{0.119} & 0.084 & 0     & 0.048 \\
0.02      & 0   & 2.35 & 0.763 & 0.044 & \resub{0.025} & 0.097 & 0.026 & 0.044 \\
\midrule
$10^{-5}$ & 0.4 & 2.35 & 0.956 & 0     & \resub{0    } & 0     & 0   & 0.044 \\ 
0.002     & 0.4 & 2.35 & 0.870 & 0.026 & \resub{0.011} & 0.002 & 0   & 0.092 \\
0.006     & 0.4 & 2.35 & 0.734 & 0.124 & \resub{0.012} & 0.012 & 0   & 0.117 \\
0.014     & 0.4 & 2.35 & 0.734 & 0.029 & \resub{0.114} & 0.040  & 0   & 0.082 \\
0.02      & 0.4 & 2.35 & 0.717 & 0.046 & \resub{0    } & 0.110  & 0.078 & 0.049 \\ 
\midrule
$10^{-5}$ & 0   & 1.9  & 0.844 & 0     & \resub{0     }& 0     & 0     & 0.156 \\
0.002     & 0   & 1.9  & 0.808 & 0.020 & \resub{0.031 }& 0.005 & 0     & 0.135 \\
0.006     & 0   & 1.9  & 0.710 & 0.069 & \resub{0.004 }& 0.073 & 0.001 & 0.142 \\
0.014     & 0   & 1.9  & 0.540 & 0.075 & \resub{0.141 }& 0.132 & 0     & 0.112 \\
0.02      & 0   & 1.9  & 0.630 & 0.050  & \resub{ 0.030}  & 0.135 & 0.045 & 0.11  \\
\midrule
$10^{-5}$ & 0.4 & 1.9  & 0.907 & 0     & \resub{0    } & 0     & 0    & 0.093 \\
0.002     & 0.4 & 1.9  & 0.759 & 0.055 & \resub{0.030}  & 0.005 & 0    & 0.151 \\
0.006     & 0.4 & 1.9  & 0.599 & 0.144 & \resub{0.029} & 0.040  & 0    & 0.189 \\
0.014     & 0.4 & 1.9  & 0.599 & 0.032 & \resub{0.138} & 0.058 & 0    & 0.173 \\
0.02      & 0.4 & 1.9  & 0.581 & 0.049 & \resub{0    } & 0.132 & 0.120 & 0.118 \\
\bottomrule
\end{supertabular}
\label{table:frac_sn_top}
\end{center}